\documentclass[aps,manuscript,overcite]{revtex}
\usepackage{graphics}

\input epsf

\def\ba{\begin{eqnarray}}
\def\ea{\end{eqnarray}}
\def\be{\begin{equation}}
\def\ee{\end{equation}}

\def\mxth{\mathsurround=0pt }
\def\xversim#1#2{\lower2.pt\vbox{\baselineskip0pt \lineskip-.2pt
    \ialign{$\mxth#1\hfil##\hfil$\crcr#2\crcr\sim\crcr}}}

\newcommand{\labeq}[1] {\label{eq:#1}}

\newcommand{\labfig}[1] {\label{fig:#1}}  

\tighten
\begin{document}
 
\begin{center}
\Large{Cosmological Perturbations\\
 in a \\
Big Crunch/Big Bang Space-time}
\end{center}

\begin{center}
Andrew J. Tolley$^1$, Neil Turok$^{1}$ 
and Paul J. Steinhardt$^{2}$\\
\vspace{.1in}
{\it
$^1$DAMTP, Centre for Mathematical Sciences, 
Wilberforce Road, \\ Cambridge CB3 0WA, UK\\
$^2$Joseph Henry Laboratories,
Princeton University,\\
Princeton, NJ 08544, USA 
}
\end{center}

\vspace{.1in}
\noindent
\begin{abstract}

A prescription is developed for matching general relativistic
perturbations across singularities of the type
encountered in the ekpyrotic and cyclic scenarios,
{\it i.e.,} a collision between orbifold planes.
We show that there exists a gauge in which the 
evolution of perturbations 
is locally identical to that
in a model space-time (compactified Milne
mod $Z_2$) where the matching of modes across 
the singularity
can be treated 
using a prescription previously introduced by
two of us.  Using this approach,
we show that long wavelength, scale-invariant, growing-mode
perturbations in the incoming state pass through the collision
and become
scale-invariant growing-mode perturbations in the expanding
hot big bang phase.

\end{abstract}
\vspace*{.1in}

\hspace*{1.in}
PACS number(s):
11.25.-w,04.50.+h, 98.80.Cq,98.80.-k
%

\section{Introduction}

The big bang singularity is one of the most
vexing puzzles in modern cosmology. Tracing time backwards,
the field equations of 
general relativity break down in an apparently irretrievable manner
some fourteen billion years ago when 
the density of matter and the curvature of space-time diverge.
Cosmic inflation does not ameliorate this disaster, but rather
tempts us to ignore it by just assuming that the universe
somehow emerged from the singularity in an inflationary state,
and that subsequent inflation washed out all of the details
of the
big bang and how inflation began.

A more fundamental point of view is that 
the singularity is a manifestation of the breakdown
of general relativity at short distances, 
which needs to be properly dealt with 
in a more consistent cosmology.
String theory and M-theory are important
suggestions as to what a more 
fundamental theory might look like, improving
on general relativity, for example, by providing
consistent perturbative S-matrices that 
include graviton processes.
If string theory is a consistent, unitary S-matrix theory,
as it is believed to be, then it is reasonable to expect that 
the cosmic singularity should be resolved within string theory,
or a 
 future development of it, in 
a satisfactory way. 
In particular,
for every `out' state 
there should be at least one `in' state.  The question 
arises: What could the `in' state have been 
which produced the hot big bang?

In recent papers, we have explored a concrete, detailed proposal for
answering this
deep question. In 
the ekpyrotic\cite{kost1} and
cyclic\cite{STu} Universe models, the origin of scale
invariant density perturbations and the
flatness,                                              
homogeneity and horizon puzzles of the standard cosmology
are all explained without
recourse to a burst of high energy primordial inflation\cite{kost1,STu}.
Instead, these puzzles are solved by physical processes occurring
prior to the hot big bang\cite{kost1,STu,phenom,gratton}, 
in a highly economical way
employing today's observed cosmological constant in an
integral manner. However, key to the  
success of these new scenarios is a consistent passage
through the big bang singularity.

At first sight,  passing safely 
through a big crunch/big bang transition
seems impossible because many physical quantities
(density, curvature) diverge there. However, in the situation
encountered in the ekpyrotic and cyclic
 brane world models, the situation is far less 
severe\cite{kost1,STu}. 
When two boundary branes 
collide, even though this {\it is} the big bang singularity
in the conventional (Einstein frame) description, 
in the background solution the density of matter and the space-time 
curvature of the branes
remain finite. Conservation of
total energy and momentum across the collision may be
consistently imposed\cite{STu}
and, once the densities of radiation and matter generated 
on the branes at the collision are fixed (by microscopic physics),
the outgoing state is uniquely determined.

However, while the background geometry describing 
a boundary brane collision 
seems to be well behaved,
it is still mathematically 
singular in the sense that one dimension disappears
at one instant of time. The space-time
ceases to be Hausdorff\cite{HE}, and since the dimensionality
of the spatial slice is only three at this moment, it is
not a good Cauchy surface.
More worryingly,  perturbations generally
diverge as one approaches the singularity,
as the result of the cosmological blue shift associated with
the collapse of the extra dimension. Nevertheless, the 
situation is more manageable than it appears to be
at first sight.
In certain gauges, the metric perturbations only diverge 
logarithmically in time\cite{ekperts}, and 
the canonical momenta associated with the perturbations
and certain other perturbation variables actually
remain finite at the singularity.

Around the brane collision, the space-time geometry may be
modeled by 
a simpler space-time 
which we shall refer to as `compactified Milne mod $Z_2$'.
This is locally flat
away from the singularity, and may
be embedded within Minkowski space-time as shown in Figure 1.
The model space-time may be thought of as describing
the collision of two tensionless $Z_2$ branes separated by
a flat bulk.
In a study of free fields on this space-time, 
two of us earlier showed\cite{Tolley} that
the construction of a unitary map between incoming
and outgoing states is not only possible but 
essentially unique.
As we review in Section II, the basic idea is to employ
normal propagation of free fields
on the Minkowski covering space-time.
This rule was shown\cite{Tolley}
to satisfy many desirable properties.  For example it defines
a vacuum two-point function of Hadamard form which is
also time reversal invariant. And in
this idealized situation with no interactions, there turns out 
to be no particle production associated with
passage through
the singularity. Some first steps were taken towards studying
interactions and these were shown to lead
to finite answers provided the coupling constant
vanishes sufficiently rapidly near the collision
event.

\begin{figure}
{\par\centering
\resizebox*{4.in}{2.7in}{\includegraphics{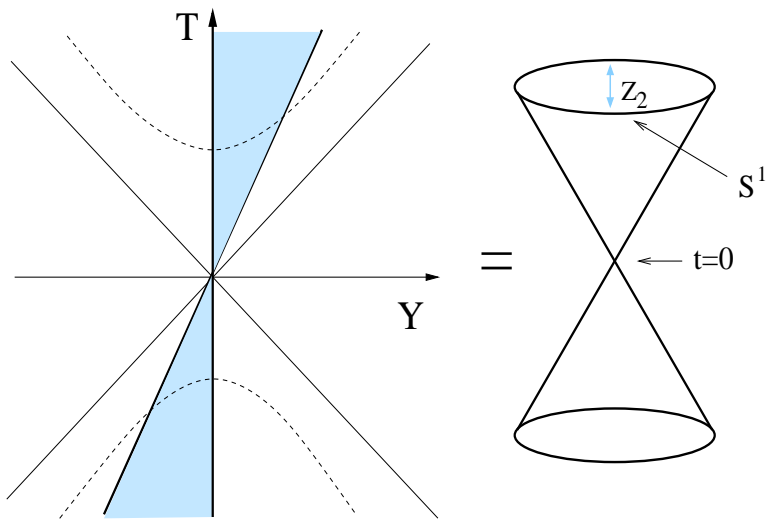}} \vskip .1in \par}
\caption{ Locally, the collision of two branes may be 
embedded in Minkowski space-time. The usual Minkowski 
space-time coordinates $T$ and $Y$ are 
are expressed as $T=t\, {\rm cosh} y$ and $Y=t\, {\rm sinh} y$,
where the Lorentz-invariant coordinate $t$ is constant on the dashed
lines. The collision event is constructed in two steps. First the
$y$ coordinate is compactified by identifying $y$ with $y+2y_0$,
to produce the double-conical space-time shown at the right.
Second, the circular sections of these cones are orbifolded by the
$Z_2$ symmetry 
$y \rightarrow 2y_0-y$. The two fixed points of
the $Z_2$ symmetry 
are two tensionless branes moving at a relative speed
of ${\rm tanh} y_0$, which collide and pass through
one another at $t=0$.
}
\labfig{boost}
\end{figure}
\noindent

The purpose of the present paper is to extend these ideas to
a study of full general relativistic perturbations 
in space-times possessing singularities of the type shown in
Figure 1.  
The usual definition of a space-time manifold is that it is a metric space
which appears locally flat. This
means that in the neighborhood of
any point $P$ it
should always be
possible to choose a coordinate system in which
(a) the metric at $P$ is the Minkowski metric, and
(b) the first derivatives of the metric with
respect to each coordinate vanish at $P$. 
The inclusion of singular points of the type shown in Figure
1 requires an extension of these rules. In particular,
the usual notion of general coordinate invariance 
becomes more subtle.
A description of the 
incoming and outgoing space-times, away 
from the singularity, should be completely independent
of coordinates since only the intrinsic geometry matters. 
However, connecting the two halves of the 
space-time across the singularity 
requires a correspondence between the `incoming' 
and `outgoing' coordinate systems.
What this means in practice is that after solving
for the metric and brane perturbations using general
relativity in the upper and lower halves (which
may be done in any gauge), one needs to choose a set of
coordinates, or gauge,
common to both halves within which
the matching is to be performed.

\begin{figure}
{\par\centering
\resizebox*{3.in}{2.7in}
{\includegraphics{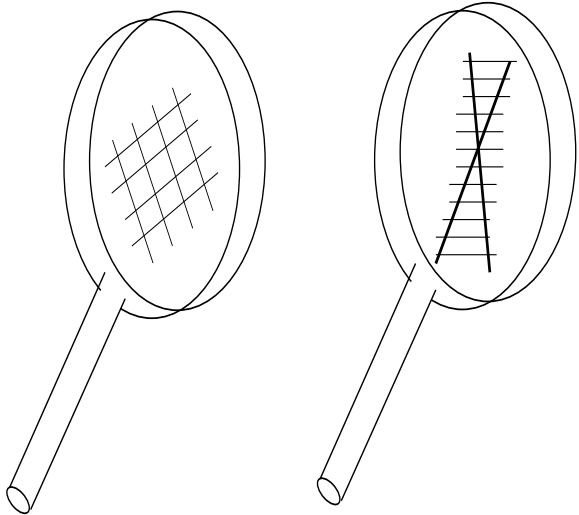}}\vskip .1in  \par}
\caption{The definition of a space-time manifold is that
when viewed `up close' (left figure), it should appear to be locally flat.
We define singular space-times
of the type we are interested in here as space-times
for which there exists a single coordinate system covering the
neighborhood of the singularity in 
both the incoming and outgoing space-times, within
which the collision event appears locally identical to
the idealized situation of tensionless $Z_2$ branes 
colliding in Minkowski space-time (right figure).
}\label{coords2}
\end{figure}

Our proposal for extending general relativity to this
type of singularity is illustrated in Figure~\ref{coords2}.
The idea is to insist that the
the upper and lower halves be connected within
a smooth set of embedding coordinates
within which the geometry
appears locally identical to that describing the model
space-time consisting of the
collision of two tensionless branes i.e. compactified Milne
mod $Z_2$.
This set of embedding coordinates, locally unique up
to Lorentz transformations,
connects the contracting
and expanding phases on either side of the bounce.
The fact that fields may  propagate across the
singularity in the model space-time shown in Figure 1
and, at the same time,
unitarity and  all the  other desirable physical
properties of massless fields propagating in ordinary Minkowski
space-time can be maintained\cite{Tolley}
makes this minimal extension of general relativity that we
propose both
reasonable and
physically sensible.

In close analogy with the definition of
a space-time manifold, we shall define `locally' by
insisting that the first two terms in a series expansion
of the metric perturbations (specifically the constant
and logarithmic terms) behave precisely as free gravitational
waves would in a compactified Milne mod $Z_2$ space-time.
The main work of the paper
will be to demonstrate that
this condition may be precisely formulated, at
least for the lowest energy modes, and that
it completely
fixes the power series expansion
in the Lorentz-invariant distance $t=\sqrt{T^2-Y^2}$
from the singularity. Within the
coordinate systems so constructed for the incoming
and outgoing space-times, we find a unique rule for
matching gravitational perturbations, in a manner
entirely analogous to
the matching of free scalar fields in the model space-time,
as
discussed in Ref.~\ref{Tolley}.

The matching procedure we propose is more subtle than that usually adopted in general
relativity. In situations where the matter stresses
change suddenly on some physically prescribed
space-like surface (for example in a phase transition),
it is normally only necessary to match the 
spatial three-metric and its normal time derivative,
without worrying about the
detailed behavior of the solutions
of the field equations.
In our case, the metric perturbations diverge
at the singularity. One might attempt to cut
the divergence off by pasting the incoming to
the outgoing space-time together on some 
arbitrary surface 
slightly away from the singularity, but it is not
known how to do this in a coordinate-invariant manner
inevitably leading to ambiguous, and usually 
 cutoff-dependent answers.
In contrast, our procedure
for massless fields including gravitational waves
on compactified Milne mod $Z_2$ 
can be formulated in terms of analytic continuation,
which is automatically coordinate-invariant, or  
in terms of a real continuation in an embedding Minkowski 
space
with asymptotically flat boundary conditions, also a coordinate-invariant
prescription.
Both methods  produce the same cutoff independent
result. Notice also that both involve global aspects 
of the space-time, and cannot be stated as a purely local
matching rule. This seems to be 
the inevitable price one has to pay
for evolving through a singularity where 
a Cauchy surface does not exist.

We have in mind of course, an application of this proposal
to the types of cosmological singularities 
encountered in ekpyrotic
and cyclic models in which two boundary branes collide as shown
in Figure 1. In particular we wish to track scale-invariant 
perturbations developed via the ekpyrotic mechanism\cite{kost1,ekperts}
in the incoming state across the
singularity and into the outgoing hot big bang phase. The conclusion
of our work is that with the prescription adopted here, 
scale-invariant, growing mode
perturbations  produced during the pre-big bang 
phase\cite{kost1,STu,gratton} 
pass through the bounce and become
scale-invariant growing mode perturbations 
in the late Universe.

Let us briefly comment on the relation
of this paper to previous studies by ourselves and others. 
Our first attempt\cite{ekperts} at matching 
perturbations across the transition was based entirely
on the study of the four-dimensional effective theory.
As we shall see, this is not sufficient to describe
the bounce, which is really five-dimensional.
Nevertheless, in that work we observed that 
certain perturbation variables, such as
the comoving energy density perturbation $\epsilon_m$ 
were finite at the singularity and could be matched
across it. The present (and far more sophisticated)
approach confirms this element of the procedure. 
The problem is that two matching conditions are needed
in  the four-dimensional effective theory
and  the first derivative
of $\epsilon_m$ turns out not to be independent of $\epsilon_m$
itself because
the differential equation is singular at $t=0$.
This leads to an ambiguity in the second matching condition.
 Based on simplicity,
we proposed matching the second derivative
and obtained an outgoing scale-invariant spectrum. However, 
we did not have any real physical justification for
this choice.

There were criticisms and alternative proposals for 
matching conditions\cite{pmps}, including
the idea that one should match the curvature
perturbation
on comoving (or constant density) 
slices\cite{lyth1,brand,jch}, a 
procedure which is 
often useful in the context
of nonsingular, expanding four-dimensional cosmology. 
In our setting, the 
comoving curvature perturbation is logarithmically
divergent at the singularity\cite{ekperts,lyth2}, but if one 
disregards this and
proceeds to match its long wavelength, constant component, this proposal
results in the growing, scale-invariant perturbations  present
in the pre-big bang phase being 
matched to a pure decaying mode in the outgoing state\cite{brand,jch}.
The result is  a complete absence of long wavelength
density perturbations in the big bang phase.
It was subsequently pointed out, however, that this null
result is atypical in the sense that, for most choices of 
matching surfaces, scale invariant
growing perturbations coming in would match to
scale invariant growing perturbations coming out\cite{dvc}.

Some of the alternative proposals are designed specifically 
for four-dimensional theories in which the
bounce from contraction to expansion 
occurs at a non-zero value of the 
scale factor\cite{peter,hbounce} (see also Ref.~\ref{gordon}). This is
accomplished by arranging for
the equation of state $w$ to violate
the null energy condition near the bounce, {\it i.e.,}
$w<-1$.  We emphasize that the ekyprotic and cyclic 
scenarios and the considerations here do not fall into 
this category. The bounce in Figure~1 corresponds
to zero scale factor in the four-dimensional effective theory and 
the four-dimensional
effective equation of state parameter is strictly positive
before the collision.

Our approach is to choose a class of gauges in which the
geometry around the collision event appears locally
identical to that describing linearized perturbations
around the model space-time, compactified Milne mod $Z_2$. 
Then we match the perturbations according to
the procedure of Ref.~\ref{Tolley} for that space-time.
An important feature of our choice of coordinates is
that the 
collision event is {\it simultaneous} in Milne time and
occurs at the background value  $t=0$
both for the incoming and outgoing state. 
That is, the limits
$t\rightarrow 0^-$ in the incoming state and 
$t\rightarrow 0^+$ in
the outgoing state correspond to the same physical
space-time surface. 

In the course of our analysis we shall uncover the 
problem with matching the curvature perturbation on 
comoving (or constant energy density) 
slices in the four-dimensional effective theory, $\zeta_4$, 
across the bounce. We shall show that $\zeta_4$ is indeed
conserved on long wavelengths both before and after the
bounce, and that furthermore on long wavelengths it
is equal to the comoving curvature 
perturbations on the branes $\zeta_\pm$. Why then
are these variables not conserved across the bounce?
The reason, detailed in 
Section V.E, is that the brane collision
event is {\it not simultaneous} in the comoving or
constant energy density time slicing. 
This is a disaster in terms of matching. In
this coordinate system, the $t\rightarrow 0^+$ and
$t\rightarrow 0^-$ space-like surfaces do not physically 
coincide and therefore perturbations should certainly
{\it not} match across them. We find that the
collision event is displaced from the 
$t=0^+$ and $t=0^-$  surfaces in these slicings by
a scale-invariant time delay, within which all
the information regarding the growing mode
perturbation is contained.
A determination of
the collision-synchronous time slices is
only possible within the full five-dimensional
theory, and our final result for the spectrum of
growing mode perturbations involves five-dimensional
parameters which cannot be re-expressed in purely
four-dimensional terms.

Distinct but closely
related are problems raised in
recent attempts to directly
study string theory on compactified Milne space-times analogous
to that shown in Figure 1.\cite{seibergetal,HP}
Since these types of background are locally flat, one can solve\cite{earlier}
the tree level field equations of string theory to all orders in
$\alpha'$, away from the singularity. It is then
tempting to calculate string scattering 
processes using a Lorentzian generalization of standard orbifold
techniques to this time-dependent case. Calculations
have been performed in analogous backgrounds, for example, 
the null orbifold and 
`null-brane' backgrounds\cite{fig,seibergetal} possessing
some remaining supersymmetry.
The result is that tree level scattering amplitudes 
develop infrared divergences which have been attributed to 
the back-reaction of the geometry near the singularity.

It is unclear what the physical 
significance of these results are yet. The breakdown of
string perturbation theory seems to indicate 
that nonlinear effects must be taken into account. But
such nonlinear effects are not necessarily disastrous for 
cosmology. For example, since the collision takes place
on a very short timescale, one
plausible possibility is that non-linearities result
in the production of microscopic 
black holes at the collision. This would be consistent
with the conclusion that perturbative
string theory breaks down, but it would be unimportant for 
cosmology. The black holes would radiate and 
decay rapidly after the bounce without having a significant effect
on the long wavelength perturbations that are relevant cosmologically.

The classical theory may provide some insight.
For example, consider classical general relativity with a 
scalar field. As the universe contracts 
towards a big crunch singularity, the gradients of the 
energy density diverge 
and one might be tempted to argue that the homogeneous 
Friedmann-Robertson-Walker (FRW) equations become invalid, 
however, this conclusion is believed to be wrong.
Instead, the 
behavior of the metric and fields becomes ultralocal.\cite{BKL}
Spatial derivatives become less important as the the universe
contracts 
and, at each point in space, the geometry 
follows a homogeneous evolution.
This occurs because, although 
the gradient terms grow, the homogeneous terms grow faster. 
A description of this subtle situation may well be 
difficult using string perturbation theory, which
relies for example upon the existence of a globally 
good gauge. However, as we shall explain in the conclusions,
there is a simple classical picture of where the nonlinearities
lead to. 
And within this picture,
we see that the nonlinear corrections would hardly alter
our final matching result.

We should also note that the string theoretic calculations
have only so far been possible in certain special models for which
the technical tools needed are available. 
In particular, they have all been done in the context of
ten dimensional string theory
at fixed coupling, 
using Lorentzian orbifolding,
with one of the {\it nine} string theory spatial dimensions
shrinking away and reappearing.  
However, this 
setup is quite different from the case proposed for the
ekpyrotic model, where the {\it tenth} spatial dimension
(of eleven dimensional supergravity), separating
the two boundary branes, was supposed to collapse and reappear. 
The eleven dimensional
theory reduces, at fixed, small brane separation, to 
string theory\cite{witten} at weak coupling. But in the
time-dependent situation we are interested in, the
coupling would actually
vanish as the branes meet. This situation
is qualitatively
different from the examples which have been studied so far.
In particular, the infinities encountered in Refs.~\ref{seibergetal}
are proportional to the string coupling. But in
the ekpyrotic model the coupling 
vanishes at the singularity.


Progress in the investigation of such singularities 
within string theory\cite{nappi} continues to 
be an active field\cite{Bala,craps,martinec,rabino,CC}.
Analytic continuation methods related to
those we employed for field theory\cite{Tolley} have been applied
to constructing string theory on similar backgrounds\cite{Tseytlin}
with less pessimistic conclusions than the
above cited works\cite{newmore}. Other approaches and methods have 
also been developed\cite{newless}.  
We have continued to develop a simpler field theoretic approach,
because it is considerably more manageable and may yield
helpful 
physical insight.
We hope that further developments of string theory
can be used to check and develop the approach presented here. 

The remainder of the paper builds in stages towards a full
calculation of the propagation of cosmological perturbations
through a bounce of the ekpyrotic/cyclic type:
\begin{itemize}
\item We first
consider
the propagation of scalar fields in a fixed background
corresponding to two
tensionless $Z_2$ branes colliding in a flat bulk as discussed in
Ref.~\ref{Tolley}.  [Section II]
\item We next consider linearized gravitational perturbations
of the same model space-time.
[Section IV]
\item Finally, we consider the full-blown calculation of cosmological
perturbations for two colliding 
branes with tension and a warped bulk.  This
calculation leads to our central result 
for the amplitude of the
scale-invariant
perturbations  propagating across the singularity into the hot big bang
phase. [Section V]
\end{itemize}

Various tools are developed along the way. Section III develops the
moduli space approximation for two colliding branes in a negative
cosmological constant bulk\cite{Randall} which we shall study
as our canonical example. We extend this
formalism, showing for example that it is exact for
empty branes at arbitrary speed and curvature. In Appendix 1
we show that the four-dimensional effective theory
consistently predicts the projected Weyl tensor
contribution to the effective Einstein equations on the
branes, and is in agreement with the recently developed
`covariant curvature' approach\cite{Shiromizu} as well as earlier metric
based approaches\cite{Wiseman,Soda}.  We also match the parameters of
four-dimensional effective theory for the homogeneous flat background
solution to the parameters of the five-dimensional theory.  
Appendix 2 discusses the  gauge invariant variables for the five-dimensional
theory
and how the position of the branes depends on the choice of gauge.
Appendix 3 works out the detailed background geometry near the bounce
in a coordinate system convenient for the perturbation calculations.
Appendix 4 concerns the choice of gauge required to have the brane collision
simultaneous at all values of the noncompact coordinates $\vec{x}$

\section{Propagation of Scalar Fields in a Collision of
 Tensionless Branes}

The idealized space-time we shall use as a model for
the singularity 
is just Minkowski space-time subject to two identifications\cite{seiberg}.
Expressing the usual Minkowski coordinates as
$T= t \,{\rm cosh} y$ and $Y=t\, {\sinh} y$, the line element
is 
\be
ds^2= -dT^2+dY^2+d\vec{x}^2 = -dt^2 +t^2 dy^2 + d\vec{x}^2.
\labeq{backg}
\ee 
The 
incoming and outgoing regions, respectively 
$t<0$ and  $t>0$, are the two halves of Milne space-time 
${\cal M}\times R^3$. We now compactify
the $y$ coordinate 
by identifying under boosts, which correspond to translations
in $y$, $y\rightarrow y+2 y_0$.
We refer to the resulting space as compactified Milne 
space-time, 
or ${\cal M}^C\times R^3$.
Finally we introduce two tensionless $Z_2$ branes
by identifying fields under reflection across the circle,
$y\rightarrow 2y_0-y$ giving the orbifolded space
${\cal M}^C/Z_2 \times R^3$, or compactified Milne mod $Z_2$.
The branes are separated by a coordinate distance $\Delta y=y_0$
which is the rapidity associated with their relative speed.
Later in the 
paper it will be convenient to choose a Lorentz frame in
which the branes are located at
equal and opposite values of $y=\pm y_0/2$.
Note that any field which is even under the $Z_2$ must obey
Neumann boundary conditions $\partial_y \varphi=0$ on
the two branes. 

The problem of propagating a free quantum field through 
a big crunch/big bang singularity of the type shown in
Figure 1
was considered in Ref. \ref{Tolley}.
The equation of motion for
a scalar field on the background (\ref{eq:backg}) is
\be
\ddot{\varphi} + {1\over t} \dot{\varphi} +{k_y^2\over t^2} \varphi
+\vec{k}^2 \varphi=0,
\label{eq:mode1}
\ee
where $k_y$ is the momentum in the $y$ direction and
$\vec{k}$ that in the uncompactified $\vec{x}$ directions.

In this paper, our main 
interest is in the lowest excitations corresponding
to the modes of the four-dimensional effective theory. In
this compactified Milne setup these modes are the
$y$-independent fields,
trivially satisfying Neumann boundary conditions on
the branes and periodicity in $y$.
For these modes, equation (\ref{eq:mode1})
is just Bessel's equation with index $\nu=0$. The two linearly
independent solutions are $J_0(kt)$ and $N_0(kt)$, behaving
for small positive $t$
as
\be
J_0(kt)\sim 1+\dots , \qquad N_0(kt) \sim
{2\over \pi}\left({\rm ln}(kt)+{\gamma}-{\rm ln}2 \right) +\dots,
\labeq{bess}
\ee
where ${\gamma}$ is Euler's constant 0.577$\dots$.
The positive (respectively negative) frequency outgoing 
modes $\psi^{(+)}$ ($\psi^{(-)}$) are those which tend
to the adiabatic 
positive (negative) frequency solutions as
$t\rightarrow \infty$. They are proportional to 
the Hankel function $H_0^{(2)}= J_0-iN_0$ (respectively 
$H_0^{(1)}= J_0+iN_0$), and converge rapidly to zero
in the lower (upper) half
complex $t$-plane. If we split the quantum field $\varphi(t,\vec{x})$
into its positive and negative frequency parts, they are well
defined respectively in the lower and upper half complex
$t$-plane. The unique analytic continuation from negative to
positive values of $t$ is then to continue the positive
frequency part below and the negative frequency above
the singularity at $t=0$.
Continuing the expressions 
(\ref{eq:bess}) around a small semicircle below $t=0$ one 
infers the relation
$H^{(2)}_0(kt)=-H_0^{(1)}(-kt)$ giving the positive frequency
mode function at negative values of $t$.
We can translate this into a matching rule
for the field $\varphi$ by writing
$\varphi= \sum a \psi^{(+)} + h.c.$, with $a$ arbitrary and complex.
The asymptotic behavior of the field $\varphi$ is then found to be
\be
\varphi \sim Q_{in} +P_{in} {\rm ln}k|t| \quad t \rightarrow 0^{-}, \qquad
\varphi \sim Q_{out} +P_{out} {\rm ln}k|t| \quad t \rightarrow 0^{+},
\labeq{asymp}
\ee
and the above continuation implies that
\be
Q_{out}=-Q_{in} +2 ({\gamma}-{\rm ln} 2)  P_{in},\qquad
P_{out}=P_{in}.
\labeq{matcha}
\ee
The  canonical momentum of the field $|t| \dot{\varphi}$ 
is actually proportional
to sign($t$)$P$. Hence,
 the field momentum
reverses at $t=0$ with this matching rule. Note, however, that the constant
term $Q$ is not preserved across $t=0$. Hence this matching rule
is not simply time reversal at $t=0$, and there is an arrow of time
across $t=0$. 

\begin{figure}
{\par\centering
\resizebox*{3.in}{3.3in}{\includegraphics{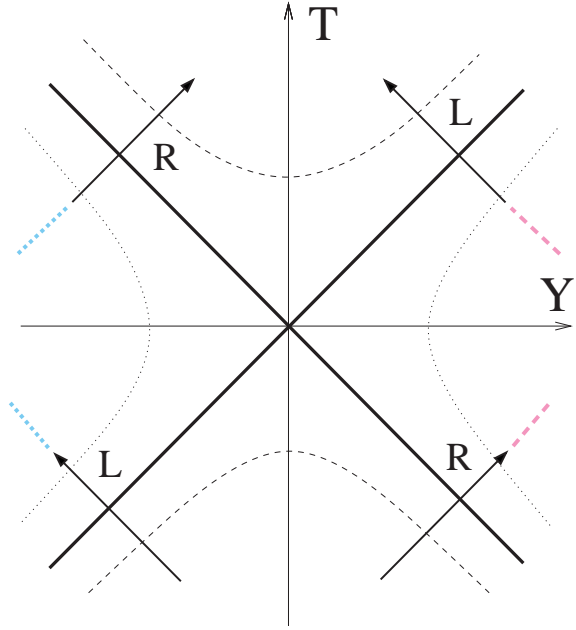}} \vskip .1in \par}
\caption{ Continuation of left and right moving modes. A free field
propagating in the lower quadrant
may be decomposed into left and right movers as it approaches
the past light cone of the origin $T=Y=0$. The left movers are
regular across  
$Y=T<0$  and may be continued into
the left quadrant
$Y<0, |T|<|Y|$. The right movers are regular across the right
segment $Y=-T>0$  and may be continued  
into the right quadrant $Y>0, |T|<|Y|$. If we impose
vanishing boundary conditions at
large Lorentz-invariant separation from the origin in the
left and right quadrants, then once we know the left mover
in the left quadrant, the right mover on the null segment
$Y=-T<0$ 
is uniquely determined, and similarly the left mover 
on $Y=T>0$.
One thereby obtains a unique matching rule
from the incoming, lower quadrant to the outgoing, upper
one.
}
\label{fig:newlrs}
\end{figure}

There is another way of looking at this rule which
is illustrated in Figure \ref{fig:newlrs}. 
Take a field configuration on one copy of the incoming wedge
and repeatedly reflect it through the boundary branes to
fill out the 
lower quadrant. The resulting configuration obeys the field
equation (even with nonlinear interactions), as long as
the equation is $Z_2$ invariant. 
The solutions to the field equation then naturally
split into left and right movers as one approaches the light cone. 
The left movers are regular on $Y=-T$ and the right movers
on $Y=T$. Each can therefore be uniquely matched across 
the appropriate segments of the past and future 
light cone of the singularity (Figure \ref{fig:newlrs}).

In this way, incoming data in the lower quadrant uniquely
determines the left moving modes entering the left quadrant and
the right moving modes entering the right quadrant. The solutions
in the left and right 
quadrants may be fully specified by choosing boundary conditions.
It is natural
to demand that the fields vanish at space-like infinity.
Once the solution in the left and right quadrants is 
determined then the left movers from the right 
quadrant and the right movers from the left quadrant may
be uniquely matched to the left and right movers
in the upper quadrant, completely determining the solution
in the outgoing state. Again, in the context of our model
spacetime compactified Milne mod $Z_2$,  this 
prescription yields exactly the same matching rule
(\ref{eq:matcha}). The advantage of this derivation 
is that it gives the clearest explanation for the sign 
change in the constant contribution $Q$, between the `in' and `out' states.
This is just due to our having imposed 
a `reflecting' boundary condition at space-like infinity.
Since in passing from the lower to the upper quadrant, 
one such reflection is involved, a relative minus sign
is acquired. And as we shall explain in Section IV, precisely
the same matching rule may be applied for gravitational
perturbations on compactified Milne mod $Z_2$. In this
case one can see that the condition of asymptotic
flatness imposed in the two unphysical quadrants
is actually coordinate invariant. 

In the case
of cosmological
interest where the branes have tension and the bulk is
warped, the sign change of $Q$ in (\ref{eq:matcha})
is still guaranteed provided two reasonable conditions 
are fulfilled. Assume that 
the low energy modes in the space-like regions (which are just the 
analytic continuation of the 
corresponding modes in the lower quadrant, obtained by setting
$t=is$ and $y= \rho-i\pi/2$, where $T=s\, {\rm sinh}\rho$ and 
$Y=s\, {\rm cosh}\rho$),
depend only on $s$ as $s\rightarrow 0$ (i.e. behave as the 
Kaluza-Klein zero modes). Second, assume that the mode selected
by the imposed boundary condition at spacelike infinity behaves,
near $s=0$, as $D+\ln(k|s|)$ with $D$ a model-dependent constant.
This is the generic behavior - for compactified Milne mod $Z_2$ we have 
$D= \gamma-{\rm ln}2$.
Then it is straightforward to show by
explicit calculation that matching the 
left/right movers across the light cone
from the lower quadrant into the left/right quadrants 
and then into the 
upper quadrant, one obtains
$P_{out}=P_{in}$
and $Q_{out}=-Q_{in} + 2 D P_{in}$. Hence we see the sign change 
of $Q$ is universal but the coefficient $D$ is not.

It is important to emphasize that all of these arguments
for the matching rule (\ref{eq:matcha}) involve the detailed 
{\it global} structure of the embedding space-time.
In particular the  $\gamma -\ln 2$ term in (\ref{eq:matcha})
is peculiar to the Minkowski embedding spacetime appropriate
for compactified Milne mod $Z_2$.
If the embedding 
space-time is warped, the corresponding constant would
be altered to some constant $D$ as explained above.
Fortunately
it shall turn out that for the case we are interested in, 
$P_{in} \ll Q_{in}$ at 
long wavelengths and hence we are insensitive to the value of $D$.
The correspondence $Q_{out}\approx -Q_{in}$ is however
universal as argued above 
and therefore reliable even in the warped case. It turns out that this sign
change is crucial in allowing scale invariant
growing perturbations to propagate across
the singularity, in the absence of radiation.
Furthermore, the sign change is interesting and
important in the nonlinear theory, as we explain in
the conclusions.

\section{The 4d Effective Theory}

In subsequent sections we shall
extend the matching rule just discussed for free scalar
fields to full general relativistic perturbations.
There are
two major complications. 
The first is the 
gauge invariance of general relativity which, as explained
above, is unusually subtle for singular 
space-times such as we are dealing with.
The second is that 
the bulk space-time is not globally Minkowski space-time but
is warped and has non-negligible $y$-dependence. 
Of course, this is related via Israel matching (see e.g. Ref.~\ref{Binetruy})
to the fact that the brane tensions
are nonzero. 

We want to solve the linearized Einstein field equations for
five-dimensional gravity coupled to a pair of colliding orbifold
($Z_2$) 
branes. For the cosmological applications, 
we need to follow the system from  times well
before the brane 
collision, when the scale-invariant
perturbations were generated, through the collision and
into the far future.
In general this would involve solving a system of coupled 
partial differential equations
in $y$ and $t$ for the bulk gravitational fields with mixed boundary
conditions following from the Israel matching conditions on the branes,
and would be well beyond an analytic treatment.

However, there is a powerful tool we can call upon
which makes the task surprisingly tractable: the moduli space 
approximation.

\subsection{The Moduli Space Approximation}

\begin{figure}
{\par\centering
\resizebox*{2.in}{2.5in}
{\includegraphics{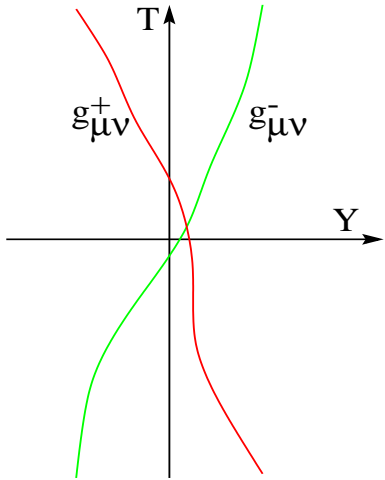}} \vskip.1in \par}
\caption{ The worldlines of the positive and negative tension
branes are plotted for some fixed value of the uncompactified
coordinates $\vec{x}$. The four-dimensional effective theory is used to
predict the intrinsic geometries of the positive and negative tension
branes, i.e. their space-time metrics $g_{\mu \nu}^+$ and $g_{\mu \nu}^-$,
according to equation (\ref{eq:branemets}).
The four-dimensional effective theory is used to
describe the incoming and outgoing perturbed branes far to the past or
future of the collision event. The brane metrics also provide
boundary data 
for the five-dimensional bulk metric
which we solve for 
as a power series expansion in time about the
collision event.
}
\label{fig:induced}
\end{figure}

On general grounds one expects the long wavelength,
low energy modes of the system to be described by a four-dimensional
effective theory, and we are only interested in low energy 
incoming states which are well described by this theory. 
We shall show that the four-dimensional
effective theory may be consistently used to predict the brane geometries
all the way to collision, thereby providing boundary data for
the bulk five-dimensional equations which we solve as an 
expansion in $t$ about the collision event. After the collision,
the four-dimensional effective theory plays an equally important
role, enabling us to track the behavior of perturbations
into the far future of the collision event (Figure \ref{fig:induced}).
The technique we describe forms the basis for our analysis 
of the singularity described in later sections, but it is
also of considerable generality and use in its own right,
since almost all of the late Universe phenomenology of
brane worlds can be most efficiently 
described using the effective theory
alone.

In this paper we concentrate on the simplest two-brane world
model consisting of 
one positive and one negative tension brane bounding 
a bulk with a negative cosmological constant $\Lambda= -6 M_5^3/L^2$
where $L$ is the AdS radius and $M_5$ the five-dimensional Planck
mass.
If the brane tensions $\sigma_\pm$
are fine tuned to the special values $\pm 6M_5^3/L$, the system allows a
two-parameter family of static solutions in which the scale factor
on each 
brane is a free parameter, or modulus. The idea of the moduli space
approach is that such parameters are promoted to space-time 
dependent fields within the four-dimensional effective theory.
In passing, we note that many 
of the methods we use in this paper should in principle extend 
to more complicated theories such as Horava-Witten theory,
in which the family of static solutions exists without
the need for fine tuning of the brane tensions.

In Khoury {\it et al.}\cite{kost1}, 
the effective action for the moduli in this system 
was computed in the low velocity
approximation, and shown to be equivalent to Einstein
gravity plus a scalar field which couples non-minimally 
to the matter on each brane (see also Ref.~\ref{ranter}).
 The derivation given here, while
more specific to the simplest brane models, is both simpler and more 
powerful. It shows that the same effective action actually
has a broader range of validity than originally anticipated, turning 
out to be exact for
empty brane configurations with cosmological symmetry,
for arbitrary spatial curvature and velocity (or expansion rate).
When matter is present,
the effective theory is 
a good approximation
as long as the density of matter is small compared to
the brane tension.
The fact that the four-dimensional effective theory
is so accurate is likely to be a special feature associated with
the lack of bulk degrees of freedom in the simplest brane world
model we are focusing on: for configurations with
cosmological symmetry, a
generalized Birkhoff theorem\cite{gregory,martinneil} holds
which guarantees that no radiation is emitted into the bulk.

Consider a positive or negative tension brane with cosmological
symmetry but which 
moves through the five-dimensional bulk. The motion 
through the warped bulk induces 
expansion or contraction of the scale factor on the
brane. As shown in Ref.~\ref{Binetruy},
the scale factor on the brane obeys a 
`modified Friedmann'
equation,
\be
H_\pm^2 = \pm {1\over 3 M_5^3 L} \rho_{\pm} 
+{\rho_\pm^2 \over 36 M_5^6} -{K\over b_\pm^2} + {{\cal C} \over b_\pm^4},
\labeq{israel}
\ee
where $\rho_{\pm}$ is the density (not including the tension)
of matter or radiation confined to the brane, $b_\pm$ is the brane
scale factor, and
$H_\pm$ is the induced Hubble constant on the positive (negative) tension brane.
We work in units such that the coefficient of the Ricci scalar in
the five-dimensional Einstein action is ${M_5^3\over 2}$.
The last term is the `dark radiation' term, where the constant
${\cal C}$ is related to the mass of the black hole in the
Schwarzchild-AdS solution discussed in Appendix 3.

We shall show that the solutions to these equations are 
precisely reproduced by a four-dimensional effective theory,
with the only 
approximation necessary being 
that the density of matter or radiation confined to the branes,
$\rho_\pm$ 
 be much smaller than the magnitudes of the brane tensions,
so that the $\rho_\pm^2$ terms in 
(\ref{eq:israel}) are negligible. 
For the particular concerns in this paper, namely the accurate
calculation of the long wavelength curvature perturbation on the branes,
it is reassuring that the four-dimensional effective theory description is
such a well-controlled approximation,
even at large brane velocities, 
in the long wavelength limit.

Choosing conformal time on each brane, and neglecting the $\rho^2$ terms 
equations (\ref{eq:israel}) become
\ba
b_+'^2 &=& +{1\over 3  M_5^3 L} \rho_+ b_+^4 -K b_+^2+ {\cal C}, \cr
b_-'^2 &=& -{1\over 3 M_5^3 L} \rho_- b_-^4 -K b_-^2 + {\cal C}.
\labeq{israela}
\ea
where prime denotes conformal time derivative. 
The corresponding
acceleration equations for $b_+''$ and $b_-''$, from which ${\cal C}$
disappears, are
derived by differentiating equations 
(\ref{eq:israela}) and using $d(\rho b^4) = b^3 (\rho-3P) \, db$, with
$P$ being the pressure of matter or radiation on the branes.
We now show that these two equations can be derived from a single
action provided we equate the conformal times on each
brane. Consider the action
\ba
{\cal S}= \int dt N d^3x\left[ -3 M_5^3 L(N^{-2}b_+'^2-{K b_+^2})  
-\rho_+ b_+^4 +3L(N^{-2}b_-'^2-{K b_+^2})-\rho_- b_-^4\right],
\labeq{newact}
\ea
where $N$ is a lapse function introduced to make the action 
time reparameterization invariant.
Varying with respect to $b_{\pm}$ and then setting $N=1$ gives the
correct acceleration equations for $b_+''$ and $b_-''$ following from
(\ref{eq:israela}). These equations are equivalent to (\ref{eq:israela}) up
to two integration constants. The constraint equation, following 
from varying with respect to $N$ and then setting $N=1$, is just the
difference of the two equations (\ref{eq:israela}) and ensures that one
combination of the integration constants is correct.  
The constant ${\cal C }$ is then seen to be just the remaining
constant of integration of the resulting system of equations and can
in effect be determined by the solutions of equations of motion
following from the action (\ref{eq:newact}).\cite{KZ}

Having shown that the modified Friedmann equations (with the neglect
of $\rho^2$ terms) follow from an action in which
${\cal C}$ does not appear, we are now able to change variables
to those in which the system appears as conventional Einstein gravity
coupled to a scalar field plus matter. 
We rewrite the action (\ref{eq:newact}) in terms of 
a four-dimensional effective scale factor $a$ and a scalar 
field $\phi$, defined by $b_+=a\,{\rm cosh}(\phi/\sqrt{6})$, $b_-=
-a\, {\rm sinh} (\phi/\sqrt{6})$. Clearly, $a$ and $\phi$ transform
as a scale factor and as a scalar field under
rescalings of the spatial coordinates $\vec{x}$. To interpret
$\phi$ more physically, note that for static branes
the bulk space-time is 
perfect Anti-de Sitter space with line element
$dY^2+ e^{2 Y/L}(-dt^2+d\vec{x}^2)$.
The separation between the branes is given by 
$d=L \ln (a_+/a_-) = L\ln \left(-{\rm coth} (\phi/\sqrt{6})\right)$,
so $d$ tends from zero to infinity as $\phi$ tends from minus infinity 
to zero.

In terms of $a$ and
$\phi$, 
the action  (\ref{eq:newact}) becomes
\be
{\cal S} = \int dt d^3x \left[-3 M_5^3 L(\dot{a}^2-K a^2) +{1\over 2} a^2 \dot{\phi}^2\right] 
+{\cal S}_{m},
\labeq{modii}
\ee
which is recognized as  the action for Einstein gravity 
with line element $a^2(t)(-dt^2+\gamma_{ij}dx^idx^j)$, $\gamma_{ij}$
being the canonical metric on $H^3$, $S^3$ or $E^3$ with curvature
$K$, and a minimally
coupled scalar field $\phi$. 
The matter action ${\cal S}_{m}$ is conventional,
except that the scale factor appearing is not the Einstein-frame 
scale factor 
but instead 
$b_+=a\,{\rm cosh}(\phi/\sqrt{6})$ and  $b_-=
-a\, {\rm sinh} (\phi/\sqrt{6})$ on the positive and negative
tension branes respectively.

Now we wish to make use of two very powerful principles. The first is
the assertion that even in the absence of symmetry,
the low energy modes of the five-dimensional theory should be
describable with a four-dimensional effective action. The second is
that since the original theory was coordinate invariant, the four
dimensional effective action must be coordinate invariant too.
Since the five-dimensional theory is local and causal, it
is reasonable to expect these properties in the four-dimensional
theory. If furthermore the relation between the four-dimensional
induced metrics on the branes
and the four-dimensional fields (i.e. the four-dimensional effective
metric and the scalar field $\phi$) is local (as one expects
for the long wavelength, low energy modes we are interested in), then covariance
plus agreement with the above results forces the relation to be
\be
g_{\mu \nu}^+ = \left({\rm cosh(\phi/\sqrt{6})}\right)^2 g_{\mu \nu}^{4d}
\qquad g_{\mu \nu}^- = \left(-{\rm sinh(\phi/\sqrt{6})}\right)^2 g_{\mu \nu}^{4d}.
\labeq{branemets}
\ee
When we couple matter to the brane metrics, these expressions
should enter the action 
for matter confined to the positive and negative tension branes respectively.
Likewise we can from (\ref{eq:modii}) and covariance 
immediately infer the effective
action for the four-dimensional theory:
\be
{\cal S} =  \int d^4x \sqrt{-g} 
\left({M_4^2\over 2 } R - {1\over 2} (\partial_{\mu} \phi)^2\right)+{\cal S}_m^-[g^-]
+{\cal S}_m^+[g^+],
\labeq{modiii}
\ee
where we have defined the effective four-dimensional Planck mass
 $M_4^2= (8 \pi G_4)^{-1} = M_5^{3} L$.

\subsection{Branes with non-zero matter density}

For most of this paper we shall only study the specially simple case of
radiation on the branes (which are 3+1 dimensional).
The matter action is then independent of $\phi$ as a result
of the conformal invariance of radiation
in 3+1
dimensions, and this will greatly simplify our analysis.
But as an aside let us for a moment consider nonrelativistic matter 
on the branes. Then there is a non-minimal coupling 
with $\phi$, leading to a source term in the
scalar field equation:
\be
\Box \phi=-\frac{1}{4}(\cosh(\frac{\phi}{\sqrt{6}})^4)_{,\phi} T^+
-\frac{1}{4}(\sinh(\frac{\phi}{\sqrt{6}})^4)_{,\phi} T^-,
\labeq{phieq}
\ee
where primes denote $\phi$ derivatives and 
the ${T^{(\pm)}}$ are the traces of the 
stress tensors for matter on the two
branes contracted with respect to the relevant brane metric.
It is interesting to see how these results compare with
what is known about brane world gravity from prior
studies\cite{GT}.
For  perfect fluids, the effective matter
Lagrangian\cite{mukhanov} reads $-\int d^4x \sqrt{-g^\pm} \rho_{\pm}$.
Hence matter on the branes couples 
to the four-dimensional (Einstein frame)
effective theory in the combination
$\rho_4=
{\cosh(\phi/\sqrt{6})}^4 \rho_++\sinh(\phi/\sqrt{6})^4 \rho_-$.
As the inter-brane distance grows, the field $\phi$ tends to zero.
Since the $\cosh$ tends to unity, we see that 
a matter source on the positive tension
brane with physical density 
$\rho_+$ contributes the same amount to the 
density seen by Einstein gravity in the four-dimensional
effective theory. Furthermore, from (\ref{eq:phieq}), the coupling
of such matter to the dilaton vanishes as $\phi$. Hence the
dilaton decouples and ordinary Einstein gravity is reproduced
in this limit. Matter on the negative tension brane behaves very
differently. If its density as seen by Einstein gravity in 
the four-dimensional
effective theory is $\rho_4$, then its physical density 
on the brane is much
larger, $\rho_- 
\sim \phi^{-4} \rho_4$, and from (\ref{eq:phieq})
it sources the dilaton field 
as $\phi^{-1} \rho_4$. Hence at small $\phi$ the 
source for the dilaton diverges and Einstein gravity is never reproduced.

The derivation we have just given of the four-dimensional effective action
starting from the modified Friedmann equations is in the present context
both simpler and more powerful than previous derivations.
It shows that the induced geometries on the branes are
correctly predicted for branes with cosmological symmetry, 
for arbitrary curvature and speed of
the branes provided only that 
that the $\rho^2$ matter terms are negligible. For these
cosmological backgrounds, 
the four-dimensional effective theory
accurately 
describes the brane collision 
even though from the Einstein frame point of 
view such a collision is highly singular in the sense that
the 4d effective scale factor 
$a$ tends to zero, the Riemann
$\phi$ tends to minus infinity in finite time.
Nevertheless, the 
brane geometries and densities 
described by $g_{\mu \nu}^\pm$ and $\rho_\pm$,
are finite and well behaved at all times. 

One surprising point about the map from five-dimensions to 
four is that the effective theory with a scalar field sourced 
by the combined energy density $\rho_4=
{\cosh(\phi/\sqrt{6})}^4 \rho_++\sinh(\phi/\sqrt{6})^4 \rho_-$ 
manages to correctly predict the solutions to the
Friedmann equations on each brane 
even though these 
are separately sourced by
$\rho_+$ and $\rho_-$. This is possible because of the integration
constants. In the four-dimensional effective 
theory the basic equations can be taken to be the
Friedmann equation (${a'}^2=\dots$) which has one integration
constant and the scalar field equation ($(a^2 \phi')'=\dots$) which
has two. So there is a three-parameter set of solutions, 
although one of these is not physical as it is just a 
rescaling of $a$. On the other
hand the two brane Friedmann
equations have two integration constants
along with the additional constant $\mathcal{C}$ which is
the dark radiation term. Consequently we have a precise match between the
integration constants showing that there is a one to one
map between the solutions of the two sets of equations. In performing
an explicit check we find that the missing information on how much
matter is contained on each brane is contained in the integration
constants for the dilaton equation.

\subsection{Relation between 4d effective theory and 5d brane parameters}

The five-dimensional background we seek to describe consists
of two parallel, flat $Z_2$-symmetric three-branes bounding
a bulk with a negative cosmological constant. In the incoming
state, as they head towards a collision, the branes are
assumed to be empty. In the 
ekpyrotic scenario, it is assumed that the brane collision
event fills them with radiation. In this section we shall
see how to describe this background setup in terms of
the four-dimensional effective theory, and in particular
we shall determine precise relations between the parameters
of the four and five-dimensional theories. 
The two brane geometries are determined according to the formulae
(\ref{eq:branemets}), and the 
background solution relevant post-collision is assumed to consist
of 
two flat, parallel 
branes with radiation densities $\rho_\pm$. The corresponding
four-dimensional effective theory has radiation density $\rho_r$,
and a massless scalar field with kinetic energy density 
$\rho_\phi$. 
It is convenient to work 
in units where the four-dimensional reduced 
Planck mass $M_4= (8\pi G)^{-{1\over 2}}$ is
unity.
The four-dimensional Friedmann equation in conformal time then reads
\be
a'^2 = {1\over 3}( \rho_r a^4 +\rho_\phi a^4) \equiv 4A_4 (r_4+ {A_4\over a^2}),
\labeq{fried}
\ee
where we have defined the constants $A_4$ and $r_4$, and used the fact that 
the massless scalar kinetic energy $\rho_\phi\propto a^{-6}$.
The reason for this choice of constants will become clear momentarily.

The solution to (\ref{eq:fried}) and the massless scalar field
equation $(a^2 \phi')'=0$ is:
\be
\labeq{sole}
a^2= 4A_4 \tau(1+r_4\tau), \qquad \phi =\sqrt{3\over 2} {\rm ln}\left( \frac{A_4 \tau}{(1+r_4\tau)}\right).
\ee
From these solutions, we reconstruct the scale factors on the branes
according to (\ref{eq:branemets}), obtaining:
\be
b_\pm= 1\pm A_4 \tau +r_4 \tau,
\labeq{branemetsol}
\ee
so we see that with the choice of normalization for the scale
factor $a$ made in (\ref{eq:fried}), 
the brane scale factors are unity at collision. For comparison,
in Ref.~\ref{STu} we parameterized the radiation density appearing in
the four dimensional effective 
theory using the  Hubble constant $H_r$ at equal density of the radiation
and scalar kinetic energy, $H_r= (2r_4)^{3\over2} / A_4^{1\over 2}$.
Also, the parameter $H_5$ used there to describe the
contraction rate of the fifth dimension may be expressed, for $r\pm L^2 <<1$
and slow velocities 
as $2A_4$.

We may now directly compare the predictions (\ref{eq:branemetsol})
with the exact five-dimensional solution given in 
in equations (\ref{eq:5dsolpars}) of Appendix 3, equating the terms
linear in $\tau$ to obtain 
\begin{eqnarray}
\nonumber & &
A_4=(1/L)(1+\frac{L^2 (r_+-r_-)}{12})\tanh(y_0/2), \\
& &
r_4=\frac{L(r_++r_-)}{12\tanh(y_0/2)},
\labeq{r4eq}
\end{eqnarray}
where $y_0$ is the rapidity associated with the relative velocity of
the branes at collision $V=\tanh(y_0)$ and $r_{\pm}$ is the value of
the radiation density $\rho_{\pm}$ on each brane at 
collision. These formulae are the exact expressions for the
four-dimensional parameters in terms of
the five-dimensional parameters neglecting contributions 
of order $\rho^2$. In fact, at leading order in $\tau$ they
are better than this since to this order the four-dimensional prediction
is exact.

For later purposes it will also be useful to define the fractional density
mismatch on the two branes as
\be
\labeq{fraction}
f=\frac{r_+-r_-}{r_+ + r_-},
\ee
so that we have
\be
r_+-r_-=\frac{12fr_4}{L} \tanh(y_0/2).
\labeq{rdef}
\ee

\subsection{Four Dimensional Perturbation Equations}

In this section, we describe the perturbations of the brane-world
system in terms of the four-dimensional effective theory. The only
cases we consider in detail are where the branes are empty or carry
radiation.
The conformal
invariance of radiation in four-dimensions greatly simplifies matters
since the scalar field then has no direct coupling to the radiation
and hence the latter evolves as a free fluid in the
four-dimensional effective theory. We elaborate on the significance of
this conformal invariance in section VI, part C.

We shall now describe the scalar perturbations,
in longitudinal (conformal Newtonian) gauge with
a spatially flat background where the scale factor and 
the scalar field are given by (\ref{eq:sole}).
The perturbed line element is
\be
ds^2= a^2(\tau)\left( -(1+2 \Phi)d\tau^2 +(1-2 \Psi\right) d\vec{x}^2 ).
\labeq{linel}
\ee
Since there are no anisotropic stresses in the linearized theory,
we have $\Phi=\Psi$ (see e.g. Ref.~\ref{mukhanov}).

A complete set of perturbation equations consists of the
radiation fluid equations, the scalar field equation of motion
and the Einstein momentum constraint:
\ba
\delta_r'&=&-{4\over 3} (k^2 v_r-3 \Phi')\cr
v_r' &=&{1\over 4} \delta_r +\Phi\cr
(\delta \phi)'' +2{\cal H} (\delta \phi)' &=& -k^2 (\delta \phi)
+4 \phi' \Phi\cr
\Phi' +{\cal H}  \Phi &=& {2\over 3} a^2 \rho_r v_r + {1\over 2} \phi'
(\delta \phi),
\labeq{perteqs}
\ea
where primes denote $\tau$ derivatives, 
$\delta_r$ is the fractional perturbation
in the radiation density, $v_r$ is the scalar 
potential for its velocity
i.e. $\vec{v}_r= \vec{\nabla} v_r$, $\delta \phi$ is the perturbation
in the scalar field, and from (\ref{eq:sole}) we have the background
quantities
${\cal H} \equiv a'/a= (1+2r_4\tau)/(2\tau(1+r_4\tau))$, and $\sqrt{2\over 3} \phi'=
1/(\tau(1+r_4\tau))$. 

We are interested in solving these equations in the long wavelength
limit, $|k\tau| \ll 1$. There are only two independent 
solutions to (\ref{eq:perteqs}), namely a growing and a decaying mode,
provided that we specify that the perturbations are 
{\it adiabatic}. Recall that the idea of adiabaticity in the
cosmological context is that for long wavelength perturbations, 
there should be nothing in the state of the matter 
to locally distinguish one region of the Universe
region from another. At each spatial location the evolution
of the densities of all the different fluids (radiation, baryons, dark
matter) should a single history in which each fluid 
evolves with the scale factor $a$ 
according to $d \rho_i = -3 (\rho_i+ P_i) \, d\ln a  =-3 \rho_i (1 +w_i) \, d\ln a $
where $\rho_i$ is its
density, $P_i$ is its pressure and $w_i$ parameterizes the equation
of state. Likewise the
total density evolves as $d \rho = -3 (\rho+ P)\,  d\ln a =-3 \rho(1+w) \, da$. 
Since the history is parameterized uniquely by the scale
factor $a$, an adiabatic perturbation can be thought of as 
arising from a fluctuation $\delta \ln a$. Hence solving all
the above equations for $\delta \ln a$, one finds
\be
{\delta_i\over (1+w_i)} \approx {\delta\over (1+w)}, \qquad i=1,\dots N,
\labeq{adi}
\ee
for adiabatic perturbations.

For the case at hand, the components of the background
energy density in the four-dimensional effective theory are scalar
kinetic energy, with $w_\phi=1$, and radiation,
with $w_r={1\over 3}$. It follows that for adiabatic
perturbations, at long wavelengths we must have
\be
\delta_\phi\approx {3\over 2} \delta_r.
\labeq{simad}
\ee
In longitudinal gauge, the fractional energy
density perturbation and the velocity potential perturbation
in the scalar field (considered as a fluid with $w=1$) are given by
\be
\delta_\phi= 2\left({(\delta \phi)'\over \phi'} -\Phi\right), \qquad
v_\phi={\delta \phi \over \phi'}.
\labeq{enpert}
\ee
From the equations (\ref{eq:perteqs}) 
above (and using $\phi' \propto a^{-2}$)
it follows that
\be
\left(\delta_\phi-{3\over 2} \delta_r\right)'=2k^2 \left(v_r-{\delta \phi \over \phi'}\right).
\labeq{velr}
\ee
Maintaining the adiabaticity condition (\ref{eq:simad}) up to
order $(k \tau)^2$ then requires that 
that the fractional
velocity perturbations for the scalar field and the radiation 
should be equal: $v_r \approx
\delta \phi/\phi'$. Expressing the radiation
velocity in terms of $\delta \phi$, the momentum
constraint (last equation in (\ref{eq:perteqs})) then yields
\be
\delta \phi \approx \left( 1+ {2\over 3} {\rho_r\over \rho_\phi}\right)^{-1}
\left({2 (\Phi'+{\cal H} \Phi) \over \phi'}\right),
\labeq{dfdr}
\ee
where $\rho_\phi= {1\over 2} \phi'^2 a^{-2}$. 

The above equations may be used to determine the leading
terms in an expansion in $|k\tau|$ of all the quantities of interest
about the singularity.
In order to compare with
Ref. \ref{ekperts}, we shall choose to parameterize the expansions 
in terms of the parameters describing 
the comoving energy density perturbation,
$\epsilon_m = -{2\over 3 } {\cal H}^{-2}  k^2 \Phi$, which 
has the following series expansion about $\tau=0$:
\ba
\epsilon_m&=& \epsilon_0 D(\tau) +\epsilon_2 E(\tau),
\labeq{sol}
\ea
where $\epsilon_0$ and $\epsilon_2$ are arbitrary constants, and
\ba
D(\tau) &=& 1 -2r_4\tau -{1\over 2} k^2 \tau^2 {\rm ln}|k\tau|+\dots
,\cr
E(\tau) &=&
\tau^2 + \dots.
\ea
For adiabatic perturbations, we obtain
\ba
\delta_\phi&=& \epsilon_0\left(-{9 \over 4 k^2 \tau^2} -{3\over 8} {\rm ln}|k\tau|
+{1\over 4} -{3\over 4} {r_4^2\over k^2}\right) +\epsilon_2 {3\over 4k^2}
+O(\tau,\tau{\rm ln}|k\tau|)\cr 
v_\phi&=& 
\epsilon_0\left({3 \over 4 k^2 \tau}(1-r_4\tau)\right)
+O(\tau,\tau{\rm ln}|k\tau|),\cr
\delta_r&=& {2\over 3} \delta_\phi +O(\tau^2,\tau^2{\rm ln}|k\tau|),\cr
v_r&=&v_\phi 
+O(\tau,\tau{\rm ln}|k\tau|),\cr
\Phi&=&\epsilon_0\left( -{3\over 8 k^2 \tau^2}+{3\over 16}{\rm ln}|k\tau|
+{15\over 8}{r_4^2\over k^2} \right) - \epsilon_2 {3\over 8k^2} 
+O(\tau,\tau{\rm ln}|k\tau|),\cr
{(\delta\phi)\over \sqrt{6}}&=& \epsilon_0\left( {3 \over 8 k^2 \tau^2}(1-2r_4\tau) 
+{1\over 16} {\rm ln}|k\tau| +
{1\over 8} + {13\over 8} {r_4^2\over k^2}\right)
-\epsilon_2 {1\over 8 k^2 }  +O(\tau,\tau{\rm ln}|k\tau|),\cr
\zeta_{4,M}&=& -\frac{1}{2k^2}\epsilon_2 +\epsilon_0 (\frac{1}{8k^2}(k^2+16
r_4^2)+\frac{1}{4} \ln|k \tau|)+
O(\tau,\tau{\rm ln}|k\tau|),
\labeq{expansions}
\ea
where $\zeta_{4,M}$ is the curvature perturbation on comoving slices
introduced by Mukhanov \cite{mukhanov}.

In an expanding Universe the adiabatic growing mode 
corresponds to a curvature perturbation, conveniently 
parameterized by $\zeta_{4,M}$. 
The decaying mode perturbation is really
a local time delay since the big bang, to which
$\zeta_{4,M}$ is insensitive but $\Phi$ is not.
As detailed in Ref.~\ref{ekperts}, in a contracting
Universe these modes switch roles so that 
the time delay mode is the growing perturbation
and the curvature perturbation is the decaying perturbation as one 
approaches the big crunch. 

The perturbations generated in the ekpyrotic/cyclic
scenarios consist of growing mode
scale-invariant perturbations in the incoming state
with no decaying mode component. These perturbations
are parametrized by $\epsilon_0/k^2$ having a scale
invariant spectrum, and since there is no decaying mode,
$\zeta_{4,M}$ is zero on long wavelengths.
After the collision,
from the four-dimensional effective theory view the
universe is expanding. Now, the growing mode perturbation
is proportional to the long wavelength part of  
$\zeta_{4,M}$.
The key question 
is whether with our five-dimensional prescription matches
the growing mode in the incoming state 
onto the growing mode in the outgoing state,
parameterized by $\zeta_{4,M}$, with nonzero amplitude.
For this to occur, the long wavelength
piece of
$\zeta_{4,M}$ must jump across the bounce. We shall see
below that this indeed occurs.

\section{Propagation of Gravitational Perturbations in a Collision of
Tensionless Branes}

In this section,
we consider the propagation of metric 
perturbations through a collision  of tensionless branes 
where the background space-time is precisely
${\cal M}^C/Z_2\times R^3$.  The analysis follows closely Section II, which
considered the propagation of
generic scalar fields in this same background. 
The results here are essential to our analysis for the physically
relevant case of colliding branes with tension (Section V) 
since our approach
is based on finding a gauge where the propagation of metric
perturbations 
through the bounce is as close as possible to the case for fixed tensionless
branes.

In this problem, it
is simplest to choose coordinates in which 
the branes remain at fixed locations and all the
fluctuations in the geometry are accounted for 
by the bulk metric perturbations. Recall that, ignoring gravity,
the background metric is 
\be
ds^2=-dt^2+t^2 dy^2 +d \vec{x}^2,
\ee
but with $y$ identified under translations  $y \rightarrow y+2 y_0$,
and the reflection 
$y \rightarrow 2 y_0-y$. The orbifold fixed
points located at $y=\pm y_0/2$ are the trajectories
of two tensionless orbifold branes. In Section II
we considered matching a scalar field across the
singularity in this space-time\cite{Tolley} and now we 
generalize the methods considered there to the case of gravitational
waves. 


A gravitational wave in
five dimensions has five independent propagating
components. If the $y$ dependence may be ignored
these five components split up in synchronous gauge
into tensor ($\delta g_{ij}$),
vector ($\delta g_{iy}$) and scalar ($\delta g_{yy}$)
components, possessing two, two and one propagating
degree of freedom respectively.
As usual in four
dimensional
cosmological perturbation theory the most interesting piece is the
scalar as this transforms nontrivially under coordinate transformations and
couples to the matter density perturbations.
The tensor pieces are especially simple
since they are trivially gauge invariant and decouple from the
matter. Finally, the vector pieces only couple to the curl component
of the matter velocities and not to the matter density
perturbation. They require
a separate analysis which will not be given here.
Furthermore, in our setup the vector modes are naturally projected out because
$\delta g_{iy}$ must be odd under the $Z_2$. Hence
the vector modes  must vanish on the
branes, and this is why there are no vector degrees of freedom in the
four-dimensional effective theory.

We shall, therefore, need only to consider the
scalar sector in what follows. The form we take for the five-dimensional
cosmological background metric is
\be
ds^2=n^2(t,y)(-dt^2+t^2 dy^2)+b^2(t,y) \delta_{ij} dx^idx^j,
\labeq{5dback}
\ee
and we write the most general scalar metric perturbation about this as
\begin{eqnarray}
	\nonumber & &
	ds^2=n^2(t,y) (-(1+2\Phi)dt^2-2Wdtdy+t^2(1-2\Gamma)dy^2  \\
	\nonumber & &
	-2 \nabla_i\alpha dx^idt+2t^2 \nabla_i \beta dydx^i) \\
	 & &
	+b^2(t,y)((1-2\Psi) \delta_{ij}-2 \nabla_i \nabla_j \chi) dx^idx^j.
\labeq{metric5}
\end{eqnarray}
For perturbations on ${\cal M}^C \times R^3$ it is straightforward
to find a gauge in which the
metric takes the form
\be
\labeq{milnegraviton}
ds^2=(1+\frac{4}{3}k^2 \chi)(-dt^2+t^2 dy^2)+((1-\frac{2}{3}k^2
\chi)\delta_{ij}+2 k_i k_j \chi )dx^idx^j,
\ee
and $\chi$ satisfies a massless scalar equation of motion on
${\cal M}^C \times R^3$.  To be precise, the gauge is
\begin{eqnarray}
\nonumber & &
\alpha=\beta=0, \qquad
\Gamma=\Phi-\Psi-k^2 \chi, \\
\nonumber & &
\Phi=\frac{2}{3} k^2 \chi, \qquad \Psi=\frac{1}{3} k^2 \chi, \\
& &
W=0.
\labeq{simppert}
\end{eqnarray}
Notice that the  non-zero variables can all be related to $\chi$
according to
\be
(\Gamma,\Phi,\Psi)=(-{2\over 3},+{2\over 3}, +{1\over 3}) \, k^2 \chi.
\labeq{ratios}
\ee
We shall, henceforth, refer to these as the `Milne ratio conditions.'
Furthermore, imposing 
the $Z_2$ symmetry, we obtain Neumann boundary conditions on $\chi$,
\be
\chi'(y_\pm)=0,
\labeq{simp}
\ee
where $y_\pm=\pm y_0/2$ are the locations of the two $Z_2$ fixed points.

In the model space-time, the lowest energy mode for
$\chi$ is $y$-independent and has 
the asymptotic form 
\be
\chi(t,y)=Q+P \ln |k t|,
\labeq{chiform}
\ee
with $Q$ and $P$ being arbitrary constants, just like the case of
scalar fields in Section II.
Our matching proposal for all the perturbation modes
is then simply
the analogue of the scalar field rule given in Section II, namely
\be
Q_{out}=-Q_{in} +2 ({\gamma}-{\rm ln} 2)  P_{in},\qquad
P_{out}=P_{in}.
\labeq{chimatch}
\ee
These relations are sufficient to determine the metric fluctuations 
after the bounce.

In later applications,
we are only interested in the long-wavelength part of the
spectrum, and, for the cases of interest,
$P$ is suppressed
by $k^2$ compared to $Q$.   
As a result, we obtain the approximate matching rule 
\be
Q_{out}=-Q_{in}, \qquad P_{out}=P_{in}.
\labeq{matchb}
\ee

The key  conditions (\ref{eq:simppert}) through (\ref{eq:simp}) are satisfied
precisely for all time in 
a  compactified Milne mod $Z_2$ background.
When tension is added to the brane and the
bulk is warped, our approach is to find a gauge which takes us as close
as possible to these conditions in the limit as $t$ tends to zero,
where the same matching rule may then
be applied.

\section{5D Cosmological Perturbations for Branes with Tension
in a Warped Background}

Our strategy for computing propagation of perturbations
when the branes are dynamical and have tension (so the bulk is warped) 
is conceptually simple: 
\begin{enumerate}
\item We use the four-dimensional effective (moduli)
theory described in Section III to provide boundary data for the
five-dimensional bulk fields. In particular, we will be interested
in the case where a nearly scale-invariant perturbations has been 
generated well before the bounce when the four-dimensional effective
theory is an excellent approximation, as occurs in ekpyrotic and
cyclic models. 
\item In the five-dimensional theory, we find a gauge which  
 approaches the Milne conditions
(\ref{eq:simppert}) through (\ref{eq:simp}) as $t \rightarrow 0$. 
In the gauge, the perturbation variables satisfy the massless scalar 
field equations of motion.
\item We use the conditions in (\ref{eq:matchb}) to propagate all
perturbation variables through the collision.
\item We match onto  the four-dimensional (moduli) theory to 
determine the cosmological results for long wavelength perturbations.
\end{enumerate}

One might worry that the  
 four-dimensional effective theory  we use to 
predict the boundary data for 
five-dimensional general relativity breaks down close to the bounce. 
However,
there are reasons to expect the effective theory remains accurate
as an approximation to general relativity
even at small times. First, in Kaluza-Klein theory, the effective
four-dimensional theory is a consistent truncation and hence
provides exact solutions of the five-dimensional theory even in
situations of strong curvature and anisotropy. 
In our case, as the branes come close, the
warp factor should become irrelevant so that the Kaluza-Klein 
picture should become more and more valid. 
Second,  
in the approach to the singularity
in general relativity\cite{damour} 
(based on the classic BKL work\cite{BKL}),
the decomposition of fields according to dimensional reduction 
does correctly predict the asymptotics of the solutions in the
limit as $t\rightarrow 0$. This suggests that the effective field
theory indeed captures the correct behavior of full five-dimensional
gravity near the singularity. 
In our detailed study of the linearized theory, we shall 
find a remarkable consistency between the predictions of
the four-dimensional effective theory near $t=0$
and the full five-dimensional cosmological perturbation
equations, and these consistency checks
are the main justification
for our use of the effective theory all the way to
the brane collision. Of course, the use of five-dimensional
general relativity near the singularity may itself be doubted 
since stringy corrections may be large there. But this
objection can only be addressed in a detailed calculation
within a string or M-theory context, which is beyond the 
scope of the present paper.

We first infer the boundary geometry
in longitudinal gauge (Section V.A) 
for which there
is a simple and precise correspondence between the four- and five-dimensional 
perturbations and both are completely gauge fixed (see also 
Appendix 2).
However, in this gauge 
the metric perturbations diverge much more rapidly 
(as $1/t^2$) than a massless scalar near $t=0$.
We shall need to transform
to a gauge where a) all the components of the metric are only
logarithmically divergent and  b) in which the 
components of the metric are in the same ratios and
obey the same boundary conditions asymptotically as $t\rightarrow 0$,
as
for the perturbed model spacetime with
two tensionless
branes in ${\cal M}^C/Z_2\times R^3$ (Section V.C).
In this gauge we can treat the 
components as massless fields and match
across the singularity as in Section II (Section V.D).

We wish to emphasize that the choice
of gauge we are making is fully five-dimensional
and is 
quite unlike that usually made in four-dimensional
cosmology for several reasons. 
In four-dimensional cosmology,
the matter present is often used to define a gauge -
for example one may choose gauges in which the total density 
or velocity perturbation is zero.
However in the five-dimensional bulk there is
never any matter present, just the cosmological 
term which is constant and,  therefore, does 
not define any preferred time-slicing. One might choose
surfaces of constant extrinsic curvature, but these
are not in any way preferred by the physics involved.
Instead, our approach focuses on the asymptotic
geometry near $t=0$, and identifying it with the
model space-time ${\cal M}/Z_2\times R^3$.
In addition to  
approximating the model space-time, it is essential
that, for the same gauge choice,
the  brane collision 
be {\it simultaneous} at all $\vec{x}$, so that the
$t=0^-$ and $t=0^+$ surfaces physically coincide. 
We shall show that our gauge choice satisfies this
latter criterion, but the standard four-dimensional gauge choices,
 {\it e.g.,} constant
density or velocity gauges,  do not.

\subsection{Longitudinal gauge moduli predictions}

In this section we wish to use the four-dimensional
effective (moduli) theory discussed in Section III to
infer the boundary data for the five
dimensional bulk perturbations.
In any four-dimensional gauge, 
the four-dimensional 
metric perturbation
$h_{\mu\nu}$ and scalar field perturbation $\delta \phi$ determine
the induced metric perturbations on the branes (in a related but
not equivalent gauge)  via the formulae
(\ref{eq:branemets}):
\be
h^{\pm}_{\mu \nu}=h_{\mu\nu}+2(\ln \Omega_{\pm})_{,\phi} \, \delta \phi \,  g_{\mu\nu},
\labeq{metpert}
\ee
where $\Omega_+= \cosh (\phi/\sqrt{6})$ and $\Omega_-
=-\sinh (\phi/\sqrt{6})$ and the metric 
perturbations are fractional
i.e. $\delta g_{\mu \nu}=a^2 h_{\mu \nu}$, $\delta g^{\pm}_{\mu
\nu}=b_{\pm}^2 h^{\pm}_{\mu \nu}$. 

This formula is particularly easy to use in five-dimensional longitudinal
gauge. (Our
definition follows that of
Ref.~\ref{Carsten}, where many useful formulae are given.)
This gauge may always be chosen, and it is completely gauge fixed
as we 
explain in Appendix 2. 
In this gauge the five-dimensional  metric takes the form
\begin{eqnarray}
	\nonumber & &
	ds^2=n^2(t,y) (-(1+2\Phi_L)dt^2-2W_Ldtdy+t^2(1-2\Gamma_L)dy^2)  \\
	& &
	+b^2(t,y)((1-2\Psi_L) \delta_{ij}) dx^idx^j, 
\labeq{5dpert}
\end{eqnarray}
Furthermore, as explained in Appendix 2,
in the absence of anisotropic
stresses the brane trajectories are unperturbed in this gauge.
An immediate consequence is that
the four-dimensional longitudinal
gauge scalar perturbation variables
$\Phi_{\pm}$ and $\Psi_{\pm}$ describing
perturbations of the {\it induced}  geometry on each brane
\be
ds^2_\pm=b_{\pm}^2 (\tau_\pm) (-(1+2 \Phi_{\pm}) d\tau_{\pm}^2+(1-2 \Psi_{\pm}) d\vec{x}^2),
\labeq{4dpert}
\ee
are precisely the boundary values of the five-dimensional
longitudinal gauge
perturbations $\Phi_\pm\equiv \Phi_L(y_\pm)$ and $\Psi_\pm\equiv
\Psi_L(y_\pm)$. Using  (\ref{eq:metpert})
and (\ref{eq:4dpert}), we find for the induced perturbations
\begin{eqnarray}
\nonumber & &
\Phi_+=\Phi_4+\frac{1}{\sqrt{6}} \tanh(\phi/\sqrt{6}) \delta \phi, \\
\nonumber & &
\Psi_+=\Phi_4-\frac{1}{\sqrt{6}} \tanh(\phi/\sqrt{6}) \delta \phi, \\
\nonumber & &
\Phi_-=\Phi_4+\frac{1}{\sqrt{6}} \coth(\phi/\sqrt{6}) \delta \phi, \\
\nonumber & &
\Psi_-=\Phi_4-\frac{1}{\sqrt{6}} \coth(\phi/\sqrt{6}) \delta \phi. \\
\labeq{pred4}
\end{eqnarray}
One subtlety in utilizing these formulas is that if $\Phi_4$ and
$\delta \phi$ are expressed as functions of four-dimensional
conformal time, then
they give the correct predictions for $\Phi_{\pm}$ and $\Psi_{\pm}$ 
on the branes in
terms of the conformal time $\tau_\pm$  on
each brane. However, when we use them as
boundary values of the five-dimensional metric 
it will be necessary to consider all the perturbation variables
as functions of 
the five-dimensional 
time $t$ entering in the background metric
 (\ref{eq:5dback}). The brane conformal times
may be expressed in terms of $t$ by integrating, 
\be
\tau_{\pm} = \int_0^t\frac{dt }{q(t,y_{\pm})},
\ee
where $q\equiv 
b/n$. So for example 
the boundary value of the bulk metric perturbation
$\Phi_L$ on the positive tension brane is given explicitly by
\begin{eqnarray}
\nonumber & &
\Phi_L(t,y_+)=\Phi_4(\int q(t,y_{+})^{-1} dt) \\
& &
+\frac{1}{\sqrt{6}} \tanh(\phi(\int q(t,y_{+})^{-1} dt)/\sqrt{6}) \delta \phi (\int q(t,y_{+})^{-1} dt),
\labeq{phil}
\end{eqnarray}
where $y_+$ is the location of the positive tension brane.
As noted, in this gauge even when we include perturbations the
branes are static and 
the Israel matching conditions 
are easily found to be
\begin{eqnarray}
\nonumber & &
\frac{b'}{b}(y_\pm)=\pm \frac{L}{6} nt\rho_{\pm}, \\
& &
\frac{q'}{q}(y_\pm)=\pm \frac{L}{2}nt(p_{\pm}+\rho_{\pm}),
\end{eqnarray}
for the background solution and
\begin{eqnarray}
\nonumber & &
\Psi_L'(y_\pm)=\frac{\dot{b}}{b}W_L \mp \frac{L}{6}nt(\delta \rho^{\pm}_L-\Gamma_L
\rho_{\pm}), \\
\nonumber & &
\Phi_L'(y_\pm)=-(\frac{\dot{n}}{n}+\frac{\partial}{\partial t}) W_L \mp \frac{L}{3}nt(\delta \rho^{\pm}_L-\Gamma_L
\rho_{\pm})  \mp \frac{L}{2}nt(\delta p^{\pm}_L-\Gamma_L
p_{\pm}), \\
& &
W_L(y_\pm) = \pm \frac{b^2 L t}{n}(p_{\pm}+\rho_{\pm}) v^{\pm}_L,
\end{eqnarray}
for the perturbations, where the right hand sides are all evaluated
at $y=y_\pm$, the locations of the positive and negative tension
branes. 
(From now on prime shall denote $\partial /\partial y$ and
dot shall denote $\partial /\partial t$.)
We can re-express $W_L$ on the branes as
\be
W_L(y_\pm)=(q^2)' v_L^{\pm},
\labeq{pertpred}
\ee
where $v_L^{\pm}$ is the longitudinal gauge
velocity perturbation of the matter on each
brane. From this one sees for example that for empty branes,
$W_L$ vanishes on the branes.

As long as the bulk matter is isotropic, as it is in our case,
the Einstein equations lead to a 
constraint which may be written
\be
G^1{}_1-G^2{}_2=0.
\ee
In longitudinal gauge this reads\cite{Carsten}
\be
\Gamma_L=\Phi_L-\Psi_L,
\labeq{5dns}
\ee
everywhere in the bulk. This is the five-dimensional analogue of the well
known four-dimensional 
no-shear condition $\Phi=\Psi$ in longitudinal gauge. 
Equation (\ref{eq:5dns}) serves to define $\Gamma_L$ on the branes in
longitudinal gauge.
Consequently we have sufficient boundary data for all the
components of the five-dimensional
 metric in this gauge. We can then perform an arbitrary five-dimensional
diffeomorphism to infer the boundary data in any gauge we choose.
Equivalently, equation (\ref{eq:5dns}) 
may be interpreted as a condition in any
gauge by using the gauge invariant variables defined in
Appendix 2. 

\subsection{Stress energy conservation}
\label{conf}

In this paper, we consider perturbations in the `in' state
which may be described as local fluctuations in a single
scalar field $\phi$ representing the inter-brane
separation. We are interested in long wavelength modes
which are completely frozen-in during the collision
event.
Hence the local processes describing the production
of radiation at the bounce
should be identical at each $\vec{x}$, and 
in the usual sense employed in cosmology, described in Section IIID, the 
perturbations should be `adiabatic'. 

As is well known, the conservation of stress energy
leads to powerful constraints on 
adiabatic density perturbations, in particular 
implying that the amplitude of the growing mode perturbation
cannot be altered on super-horizon scales. 
In this section we discuss this constraint
and show how it implies the 
spatial curvature of comoving (or constant energy density) slices 
is conserved on large scales both 
for the brane geometries and for the
four-dimensional effective theory.
We shall restrict ourselves to 
considering only radiation on each brane. 
This considerably simplifies the analysis because
when the matter on each brane is conformally invariant,
as explained above, in the four-dimensional effective theory
the scalar field decouples from the matter and can be 
treated as an independent fluid.

First we need to generalize the usual notion of adiabaticity
to deal with perturbations in the radiation densities on each brane.
As mentioned above, radiation couples to the scale factor
$\Omega_+(\phi) a$ on the positive tension brane and $\Omega_-(\phi) a$
on the negative tension brane with notation as in the 
previous section. Conservation of the radiation density on each brane
reduces at long wavelengths 
to $d\rho_\pm = -4 \rho_\pm\, d \ln a -4 \rho_\pm\, d \ln \Omega_\pm $.
Likewise we have for the radiation density in the four
dimensional effective theory $d\rho_4 = -4 \rho_4 d \ln a $.
Hence solving for $\delta \ln a$ as in 
Section IIID, we infer the adiabaticity condition for radiation 
on the branes to be 
\be
\labeq{deltapm}
\delta_{\pm} = \delta_4 -4 (\ln \Omega_{\pm})_{,\phi} \delta \phi.
\ee
The equation for conservation of energy in four dimensions
can be written 
in the form\cite{Carsten}
\be
\dot{\zeta_B}=\frac{1}{3} k^2 v_L,
\labeq{consz}
\ee
where 
\be
\labeq{zetab}
\zeta_B=\Psi-\frac{1}{3(1+w)}
\delta,
\ee
is the gauge-invariant variable measuring the spatial curvature
perturbation on constant density hypersurfaces, as originally 
defined by Bardeen\cite{Bardeenchina}. The quantity $v_L$ is
the gauge invariant scalar velocity potential, equal to
the velocity potential in longitudinal gauge 
(so that $\nabla_i v_L$ is the scalar part of the velocity
perturbation). 

At long wavelengths $k\rightarrow 0$, equation 
(\ref{eq:consz})
implies that $\zeta_B$ is conserved, provided the velocity
perturbation does not grow with scale. This property
is very powerful since it means that under most 
circumstances, as long as modes remain outside the horizon 
$\zeta_B$ can be trivially extrapolated from the early to the late
Universe, where it gives the amplitude of the growing mode
adiabatic density perturbation, the main quantity of observational
interest today.

The above definition (\ref{eq:zetab}) applies equally
on each brane and in the four-dimensional effective theory,
provided the terms on the right hand side are 
appropriately interpreted. On the branes, we have
\be
\zeta_{B,\pm}=\Psi_\pm-\frac{1}{4} \delta_\pm,
\labeq{zetabb}
\ee
where $\delta_\pm$ are the fractional perturbations
in the radiation densities on each brane, and
$\Psi_\pm$ is the perturbation in the brane spatial
metric. Using (\ref{eq:pred4}), written as
\be
\Psi_\pm=\Psi_4-(\ln \Omega_{\pm})_{,\phi} \delta \phi,
\labeq{psibra}
\ee
and the adiabaticity
condition (\ref{eq:deltapm}) we see that 
the four-dimensional effective value of Bardeen's variable,
$\zeta_{B,4}\equiv \Phi_4-\frac{1}{4}\delta_4$ is
in fact identical to $\zeta_{B,\pm}$ on long wavelengths.

Our final result will in fact more naturally
emerge in terms of another gauge invariant variable, the 
curvature perturbation on comoving
slices, emphasized by Mukhanov and others\cite{mukhanov}.
This is defined as 
\be 
\zeta_M=\Psi+{\cal H} v,
\labeq{zetam}
\ee
with $v$ the velocity potential
and ${\cal H}\equiv d\ln a(\tau)/d\tau$ the conformal
Hubble constant. Again this may be interpreted on
either brane or in the four-dimensional effective theory.
But adiabaticity requires that
the fluid velocities be identical on long wavelengths
for each fluid component. Therefore we must have
$v_\pm =v_4=\delta \phi/\phi' $ (from (\ref{eq:enpert}).
This is also seen to be consistent with (\ref{eq:consz})
and the equality of the Bardeen variables 
$\zeta_{B,\pm}=\zeta_{B,4}$ which we have just shown.

The scale factors on each brane are 
related to the four-dimensional effective scale factor
via 
$b_\pm= \Omega_\pm a$. Recalling that the conformal 
times on the branes are the same as that in the effective
theory, we have ${\cal H}_4={\cal H}_\pm 
-(\ln \Omega_{\pm})_{,\phi}\phi_{,\tau}$. Using 
$v_4=v_\pm= v_\phi=\delta \phi/(\phi_{,\tau})$ we find
\ba
\zeta_{M,4}&\equiv& \Psi_4+{\cal H}_4 v_4
= \Psi_4+{\cal H}_\pm v_\pm -(\ln \Omega_{\pm})_{,\phi}\delta \phi\cr
&=& \Psi_\pm+{\cal H}_\pm v_\phi,
\labeq{zetameq}
\ea
which is just $\zeta_{M,\pm}$.
So for adiabatic perturbations and
at long wavelengths, the comoving curvature perturbations
on the branes are both equal to that in the four-dimensional
effective theory. As is well known, the 
latter is conserved for 
for adiabatic perturbations at long wavelengths. It
follows that away from the bounce, 
$\zeta_{M,\pm}$ are both conserved as
well.
As we discussed in the introduction, and will
detail below, this does {\it not} imply they are
conserved across the bounce.

We will use (\ref{eq:zetameq}) below, but we should point
out one minor subtlety. We shall be performing all
our calculations in five-dimensional time $t$, not
four-dimensional conformal time. The velocity $v_\phi$
is not a scalar under coordinate transformations,
and we shall need to multiply $v_\phi$ by a factor of
$q$ when we re-interpret equation (\ref{eq:zetameq}) in
terms of the five-dimensional time $t$. 


\subsection{Transformation  to  Milne gauge}

Our philosophy  is to evolve cosmological  perturbations 
through the bounce in a 
`Milne gauge' where they
behave as closely as possible
to  gravitational waves on
${\cal M}^C/Z_2 \times R^3$, as described in Section IV.  Then,
we can use the same matching conditions (\ref{eq:matchb}) 
to determine the perturbation spectrum after the bounce. 

The Milne gauge we use is chosen to 
match the gauge choice (\ref{eq:simppert})
in Section IV  up to corrections of 
order  $t$ and $t \, {\rm ln} \, |kt|$ due
to
the finite brane tension, radiation densities and the 
warp factor. We still have enough coordinate freedom to
set 
three  linear combinations 
of the metric perturbations equal to zero for all $t$,
and we 
choose
\be
\labeq{gauge}
\alpha=\beta=0, \qquad \Gamma=\Phi-\Psi-k^2 \chi.
\ee
A remarkable feature of this choice is that
the constraint equation (\ref{eq:5dns}) implies
that $\chi$ obeys the equation for a massless
scalar field on the unperturbed background for all times.
From  (\ref{eq:gauge})
and (\ref{eq:gages}) in Appendix 2, we find
\be
\nabla^2 \chi=-\frac{1}{t}
 \frac{\partial}{\partial t}(t\frac{\partial \chi}{\partial
t})+\frac{1}{t^2} \frac{\partial^2 \chi}{\partial y^2} -3 {\dot{b}\over b}
\frac{\partial \chi}{\partial t} +\frac{3}{t^2} {b'\over b} \frac{\partial
\chi}{\partial y} -\frac{k^2b^2}{n^2} \chi=0.
\labeq{eom}
\ee
This result is remarkable in that it is independent of the precise
details of the background bulk geometry and the form of the stress energy in
the bulk, assuming only that no anisotropic stresses are present.
 
The remaining gauge freedom is 
of the form $x^{\mu} \rightarrow x^{\mu}
+\xi^{\mu}$ where
\be
\xi^t=\frac{b^2}{n^2} \dot{\xi}^s , \qquad \xi^y=-\frac{b^2}{n^2 t^2} {\xi}'^s,
\labeq{g1}
\ee
provided that $\xi^s$ also satisfies a massless scalar field equation
\be
\nabla^2 \xi^s=0.
\labeq{g2}
\ee
Since $\chi$ transforms as $\chi \rightarrow \chi
+\xi^s$, and $\chi$ is zero in longitudinal gauge, it follows that 
$\chi$ in the gauge we use is, in fact,  precisely value of the
spatial coordinate transformation $\xi^s$ needed to get to a
gauge satisfying (\ref{eq:gauge}) from five-dimensional longitudinal gauge. Furthermore,
$\xi^t$ and $\xi^y$ may be inferred from $\chi=\xi^s$ via (\ref{eq:g1}).

To completely fix the gauge within the family specified by
(\ref{eq:gauge}), 
we need to specify boundary conditions 
for the field $\chi$ on the two branes, and initial conditions
on some space-like surface. As a first guess, one might 
consider choosing to 
fix the gauge by specifying Neumann boundary conditions
on the branes (i.e. $\chi'(t,y_\pm)=0$)
for all time, as in Section IV.
One can easily prove
that in this gauge, as in longitudinal gauge, the brane
trajectories are unperturbed. This follows from 
the formula (\ref{eq:g1}) upon setting $\xi^s=\chi$ as 
noted above. This is very important: 
it follows that in this Neumann gauge the brane collision is simultaneous 
and occurs at precisely $t=0$ for all $\vec{x}$. Furthermore the 
Neumann gauge $\chi'(t,y_\pm)=0$ for all $t$ is a good gauge in
the sense that none of the metric components diverge worse 
than logarithmically. 

However, it turns out that setting
$\chi'(t,y_\pm)=0$
for all time is too strong a condition. One cannot
choose Neumann gauge for all time and also have 
\begin{eqnarray}
W&=&0 \, + \, {\cal O}(t, \, t \, {\rm ln} \, |kt|) \\
\Phi&=&\frac{2}{3} k^2 \chi \, + \, {\cal O}(t, \, t \, {\rm ln} \, |kt|)  \\
\Psi&=&\frac{1}{3} k^2 \chi \, + \, {\cal O}(t, \, t \, {\rm ln} \, |kt|),
\labeq{gaugez}
\end{eqnarray}
consistent with the behavior in the model space-time 
(\ref{eq:simppert}) at leading order in $t$ and $t \, {\rm ln} \, |kt|$.
The resolution is simple: 
we need to perform a small gauge transformation away 
from Neumann gauge in which we maintain only the asymptotic
vanishing of the proper normal derivative of $\chi$
as $t$ tends to zero, i.e. we impose that
\begin{equation}
n^{-1} t^{-1} \chi'(y_\pm)= 0 \, + \, {\cal O}(t, t \, {\rm ln} \, |kt|),
\labeq{normal}
\end{equation}
on the two branes.
With this choice we are able to impose all of the conditions 
in (\ref{eq:gaugez}) as well as 
(\ref{eq:gauge}).
This
small gauge transformation away from Neumann gauge shifts the locations
of the branes, $y_\pm$, but only by a finite amount.
As discussed in Appendix 4, this means that the rapidities
of the branes are perturbed in our chosen
gauge, but the collision event is still simultaneous. 

Our reason for expecting that we can choose a gauge specified
by (\ref{eq:gaugez}) and (\ref{eq:normal}) is that
when the branes approach
the warp factor should become increasingly irrelevant 
and the real background space-time should asymptotically
approach the model space-time ${\cal M}/Z_2 \times R^3$.
We expect the low energy modes we are interested in 
to behave as the lowest
Kaluza-Klein modes in this limit, i.e. becoming independent
of $y$. 
Within the
class of gauges specified by (\ref{eq:gauge}), we shall
indeed see that there are solutions for the perturbations 
in which 
all the perturbation components behave like $Q +P {\rm ln}|kt|$
as $t$ tends to zero.
The Milne ratio condition (\ref{eq:ratios}) turns out to be 
automatically satisfied by the coefficients of the logarithms. 
Fixing the constant terms to be in the Milne ratios further 
fixes the 
gauge up to a residual two-parameter family and 
imposing asympotically Neumann boundary conditions (\ref{eq:normal})
on both branes then completely fixes the gauge.

Imposing asymptotically Neumann boundary conditions turns
out to have various other natural consequences. For example
in this gauge, all the metric perturbation components 
possess identical asymptotic behavior (i.e. constant
and logarithmic terms) on the two branes, 
as $t$ tends to zero, consistent with their behavior
as a lowest Kaluza-Klein mode. Furthermore, 
there is a simple geometrical consequence of this choice
which we explain in Appendix 4, namely that the
in this gauge the perturbations to the embedding $(T,Y)$ 
coordinates of the brane collision event actually vanish so
the branes collide at precisely the background values
of $T$ and $Y$.

The
non-zero perturbations in our chosen class of 
gauges are $\Phi$, $\Psi$, $W$ and
$\chi$ along with $\Gamma$ which is fixed by the gauge choice
(\ref{eq:gauge}). 
All  the gauge freedom is contained in the solution for
$\chi$. To see this we note that if we know the solutions for $\chi$
we can immediately infer $\Phi$, $\Psi$ and $W$ from the values in
longitudinal gauge via the formulae from Appendix 2, (\ref{eq:gages}),
which with (\ref{eq:gauge}) imply 
\begin{eqnarray}
\labeq{transf}
	\nonumber & &
	\Phi=\Phi_L-\dot{(q^2 \dot{\chi})}-\frac{\dot{n}}{n}q^2
	\dot{\chi}+\frac{n'}{n}(\frac{q^2 \chi'}{t^2}), \\
	\nonumber & &
	W=W_L-(q^2 \dot{\chi})'-t^2 \dot{(\frac{q^2 \chi'}{t^2})}, \\
  	& &
	\Psi=\Psi_L+\frac{\dot{b}}{b}q^2 \dot{\chi}
	-\frac{b'}{b} \frac{q^2 \chi'}{t^2}.
\end{eqnarray}
Here as above, $q\equiv b/n$.

Our goal then is simply to determine $\chi$ to sufficient order in $t$ to be 
able to compute all the other components from (\ref{eq:transf}). 
As we have already
explained, in our chosen class of gauges $\chi$ satisfies the massless scalar
equation (\ref{eq:eom})
at all times. To specify a complete solution we need to specify both
Cauchy data on some constant $t$ hypersurface between 
the two brane worldsheets, plus boundary conditions on the
two branes. The boundary data will be obtained from the 
four-dimensional effective theory, and we 
 make the conjecture
that the bulk solution which is consistent with these
data 
will behave near $t=0$ like a
Kaluza-Klein zero mode on ${\cal M}^C/Z_2\times R^3$, which is to say that
the perturbations should be independent of $y$ as $t$ tends to zero.
In practice this means we will look for a solution which is
asymptotically of the form $\chi=Q+P \ln |kt|$, independent of
$y$. 
This assumption formally provides
the Cauchy data once we determine $Q$ and $P$ (see below).

At higher orders in $t$, we shall allow
for arbitrary Neumann boundary conditions, which we shall parameterize as
\be
\labeq{bc}
\chi'(y_\pm) = {1\over 2} a_2^{\pm}t^2+O(t^3,t^3{\rm ln} |kt|).
\labeq{defgaugea}
\ee
As 
 explained above, we shall 
adjust the coefficients $a_2^\pm$ to obtain the correct Milne ratios.
Note that there can be no $O(1)$ term since we are assuming that
$\chi$ is asymptotically of the form $\chi=Q+P \ln t$,
independent of $y$, and the $O(t)$ term is prohibited
by our condition (\ref{eq:normal}). In principle we
could also include $t \ln t$ and $t^2 \ln t$ terms but we shall find
that the Milne ratio conditions (\ref{eq:ratios}) 
are sufficient to rule these terms
out. 

The form of the series expansion
for $\chi$, implied by its equation of motion (\ref{eq:eom}), is
\begin{eqnarray}
\nonumber & &
\chi(t,y)=(Q+(f_1(y)+c_1 \cosh y+c_2 \sinh y) t +f_2(y) t^2/2 +O(t^3)) \\
& &
+P
\ln |k t| (1-\frac{1}{4}k^2 t^2 +O(t^3)),
\labeq{chiseries}
\end{eqnarray}
where $f_1(y)$ and $f_2(y)$ are two functions of $y$ that are
obtained as solutions of second order differential equations in $y$ with
boundary conditions derived from (\ref{eq:bc}). 
We choose to define $f_1$ so that 
$f_1'(y_{\pm})=0$. Therefore if $\chi$ satisfies the asymptotically
Neumann condition (\ref{eq:normal}) on both branes, we must have
$c_1=c_2=0$. A geometrical interpretation of this condition
is explained in Appendix 4.

Using the expressions for $b(t,y)$ given 
in Appendix 3, equation (\ref{eq:5dsolpars}), in the equation of
motion (\ref{eq:eom}), for $\chi$, we find at order $t^{-1}$
the following differential equation must be satisfied by $f_1(y)$:
\be
f_1''-f_1-{P\over 2 L\sinh y_0} \left((6+r_+L^2) \cosh (y+{y_0\over 2})
-(6-r_-L^2) \cosh (y-{y_0\over 2})\right)=0,
\labeq{feq}
\ee
A similar equation for $f_2$ is found at order 
$t^0$. The solutions are messy
in general but simpler when no radiation is present,
for example in this case we have
\be
f_1(y)={3 P \over 2 L \cosh(y_0/2)} \left(y \cosh y - (1+{y_0\over 2} 
\tanh {y_0\over 2})\sinh y \right).
\labeq{fsol}
\ee

By substituting (\ref{eq:chiseries}) into (\ref{eq:eom}) and 
imposing the boundary conditions (\ref{eq:bc}) at each order,
the solution for $\chi$ up to 
$t^3$ corrections is completely determined 
in terms of the four constants in total: $Q,P$,
and $a_2^{\pm}$.  From this solution for $\chi$, 
equations (\ref{eq:transf}) then determine
all the other components of the metric perturbations at leading order in $t$, on
each brane. 

Let us start by determining the spatial curvature perturbation
$\Psi$ on each brane. From (\ref{eq:zetameq}) and  (\ref{eq:transf})
we find 
\be
\Psi=\zeta_{4,M}+q \frac{\dot{b}}{b}(q \dot{\chi}-v_{\phi})
-\frac{b'}{b}\frac{q^2 \chi'}{t^2}.
\labeq{important}
\ee
We require that $\Psi$ be only logarithmically divergent.
Since from (\ref{eq:expansions}) we 
have that $v_{\phi}=3\epsilon_0/(4k^2
\tau)+O(1)$, diverging as $t^{-1}$ as $t \rightarrow 0$, we see
from (\ref{eq:chiseries}) and the expressions for the background
metric functions in Appendix 3 that only 
$\dot{\chi}$ can cancel that divergence, which requires that
\be
P=\frac{3\epsilon_0}{4k^2}.
\labeq{peq}
\ee
This condition ensures that the curvature $\Psi$ 
in our gauge and the 
comoving curvature $\zeta_{4,M}$ in the four dimensional effective theory
only differ by a constant at leading order in $t$. However, it
shall be very important that the constant is nonzero.
As we shall see, the constant
represents the time delay between the two time-slicings, 
and it is the key to why $\zeta_{4,M}$ jumps
across the singularity.

We shall now show that it is possible to choose the three
remaining gauge constants 
$Q$ and
$a_2^\pm$ so that the metric takes the 
canonical Milne gauge form asymptotically as $t$ tends to zero.
First, in this gauge all the metric perturbations behave as
$Q +P{\rm ln}|kt|$, as $t$ tends to zero, but with different 
constants $Q$ and $P$ for each component.
Substituting (\ref{eq:chiseries}) and (\ref{eq:peq}) 
into (\ref{eq:transf}), with $\Psi_L$ given from (\ref{eq:zetameq}),
$\Phi_L$ given from (\ref{eq:phil}) and $W_L$ given from
(\ref{eq:pertpred}), one finds that the logarithmic terms 
are actually all in the correct Milne ratios (\ref{eq:ratios}),
and also that $W$ vanishes to leading order, independently
of the undetermined constants.
Furthermore, the logarithmic terms obey
$\Phi(y_+)-\Phi(y_-)=0$ and $\Psi(y_+)-\Psi(y_-)=0$,
consistent with our assumption that the Kaluza Klein
zero mode dominates.

The gauge constants $Q$, $a_1^\pm$ and
$a_2^\pm$ do,  however, affect the $t$-independent 
constant terms in each 
metric perturbation component. Two of the constants
are fixed once one sets
the constant terms in $\Phi$ and $\Psi$ to 
their Milne ratio values $(2/3)k^2 \chi$ and
$(1/3)k^2 \chi$. 
We also want to
ensure that all components of the metric perturbations
behave asymptotically like a Kaluza-Klein zero-mode, becoming
independent of $y$ as $t$ tends to zero. We check this 
by comparing the values of $\Phi$, $\Psi$ and $W$ on the 
two branes. 
The difference of $\Psi$ on the two branes turns out to be
independent of the choice of the gauge constants as
$t$ tends to zero, 
\be
\Psi(y_+)-\Psi(y_-)=O(\rho_{\pm}^2L^2).
\ee
Since the moduli space approximation was derived neglecting $\rho_{\pm}^2 L^4$
corrections, to the  
order we can trust the calculation,
$\Psi$ is equal at the two brane locations. The difference
of $\Phi$ on the two branes is not automatically zero at leading order
however, and setting it zero provides the  additional equation
needed to
determine the third constant.

The result of these calculations is that the solution for $\chi$ up to
$O(t^3)$ and the leading order behavior of the other components of
the metric is completely determined.
Explicitly we find\cite{mathematica}
\begin{eqnarray}
\labeq{match}
\nonumber
\Psi(t)=& & \zeta_{4,M}(t) + \frac{\epsilon_0 \tanh(y_0/2)}{32k^2L^2
\cosh^2(y_0/2)}\left(18(y_0-\sinh(y_0))-L^2 (r_+-r_-) (-3 y_0+\sinh(y_0)\right)
\\ & &
+O(\rho_{\pm}^2 L^2,t,t\ln|kt|).
\end{eqnarray}
Since $\Psi$ is one of the variables which we match in our chosen gauge,
it follows that our prescription is quite different
to matching the comoving curvature perturbation $\zeta_{4,M}$
four-dimensional effective theory. As we shall explain,
the additional terms
in (\ref{eq:match}) allow the propagation of growing
mode perturbations across the singularity.

\subsection{Matching Proposal}

The requirement that around the collision event 
the geometry looks locally like ${\cal M}^C/Z_2 \times R^3$
has completely fixed the gauge in the incoming and
outgoing states. As elaborated in Appendix 4, the 
asymptotically Neumann boundary condition (\ref{eq:normal})
further ensures that the collision event is simultaneous
in our gauge, an essential property for matching perturbations
since the space-like surfaces defined by $t\rightarrow 0^+$ and
$t\rightarrow 0^-$ then physically coincide. 

Furthermore as we have discussed 
this gauge is special in that the induced geometry on each
brane is asymptotically the same at collision. In general 
if a brane is moving, the
values of the bulk perturbations $\Phi$ and $\Psi$ evaluated on the
branes differ from the induced values $\Phi_{\pm}$ and
$\Psi_{\pm}$. The differences are given by 
\begin{eqnarray}
\nonumber & &
\Phi_{\pm}-\Phi(y_{\pm})= -\frac{b^2}{n^2 t^2}  \frac{n'}{n}\chi' \\
\nonumber & &
\Psi_{\pm}-\Psi(y_{\pm})= +\frac{b^2}{n^2 t^2} \frac{b'}{b}\chi'.
\end{eqnarray}
Since $n'/n \propto t$ and  $b'/b \propto t$ as $t \rightarrow 0$, 
in the presence of matter on the branes, 
if we make the requirement that the
metrics on each brane are asymptotically identical this fixes
$\chi'=0+O(t^2)$, which is what we have required.
Physically this seems
a natural choice of 
gauge because when two ordinary branes collide, 
the induced geometries are identical at the collision moment.
This interpretation is also consistent
with the predictions from the four-dimensional effective theory 
where the brane metrics are give by
$g^+_{\mu\nu}=(\cosh(\phi/\sqrt{6}))^2 g_{\mu\nu}$ and
$g^-_{\mu\nu}=(-\sinh(\phi/\sqrt{6}))^2 g_{\mu\nu}$. Since the
collision 
corresponds to $\phi \rightarrow -\infty$, in the limit
there is no difference between the two conformal factors, and
the brane geometries appear identical.

So in the gauge we have fixed by the requirement that
the perturbations behave asymptotically like those on
the model space-time ${\cal M}/Z_2\times R^3$, 
the collision event 
is synchronous, the Milne ratio conditions are satisfied, 
the boundary conditions are asymptotically 
 Neumann, and the geometries on each brane are
asymptotically the same both before and after the collision.
Our 
matching proposal amounts to relating the
geometry on these chosen time slices across the collision.
We believe that these are sufficiently desirable properties to
justify this as the natural gauge in which to perform the matching,
and from now on we shall take this to be our complete gauge fixed
matching gauge.

Let us now return to our final formula (\ref{eq:match}) 
to infer its meaning in the context of ekpyrotic and
cyclic models. In those scenarios\cite{ekperts}
the quantity $\epsilon_0/k^2$ has an 
approximately scale-invariant long wavelength spectrum in the 
incoming state.
The first point to make 
is that even in an in-state with no radiation present the dimensionless 
curvature
perturbation on spatial slices $\Psi$ in our gauge has a scale
invariant spectrum, since
\be
\Psi=\zeta_{4,M}+\frac{9 \epsilon_0 \tanh(y_0/2)}{16 k^2L^2
\cosh^2(y_0/2)}\left(y_0-\sinh(y_0)\right).
\ee 
Recall that $y_0$ is the relative rapidity and 
$V_{in} \equiv  \tanh(y_0)$ is
the incoming relative velocity between the two branes.
Then, at small velocities this gives
\be
\Psi=\zeta_{4,M}-\frac{3}{64}\frac{\epsilon_0}{k^2L^2}V_{in}^4.
\labeq{sec}
\ee
We may interpret this geometrically as follows.
In the absence of radiation there is no real meaning to
the curvature perturbation on the branes but if we imagine that there
is a small density of radiation coming in, and the perturbations
are adiabatic, we can infer the comoving curvature perturbation
on the brane, $\zeta_{\pm,M}$, so (\ref{eq:sec}) becomes
for long wavelengths
\be
\Psi_{\pm}=\zeta_{\pm,M}-\frac{3}{64}\frac{\epsilon_0}{k^2L^2}V_{in}^4.
\ee
Since $\zeta_{\pm,M}$ and $\Psi$ are the spatial curvature perturbations 
of the
branes as respectively measured in the comoving timeslicing and
in our chosen timeslicing (in which the collision is at $t=0$),
it must be that 
the additional piece arises from a time translation between the two
gauges. That this is so is verified when one traces back the origin of this
term to the second term in equation (\ref{eq:important}). 
As explained in the introduction, comoving
gauge (or equivalent constant energy density gauge) are
bad gauges to match in because the brane collision is
not simultaneous in those gauges. Since our prescription
is to propagate $\Psi$ across the collision, 
the jump in $\zeta_\pm$ is due to the time
delay occurring between the collision-synchronous surfaces
in our gauge, and those of the comoving/constant density
surfaces. The key to our result is that 
in the comoving or constant density 
gauge the time delay between $t=0$ and the 
actual brane collision event
has a scale invariant
spectrum.

In fact, using (\ref{eq:expansions}) and (\ref{eq:match}) to find
$Q$ and $P$ before and after the bounce for all components of the 
metric perturbations and matching 
 according to
the rule given in equation (\ref{eq:matchb}) results
in $\zeta_{4,M}$ inheriting {\it two}
separate scale-invariant long wavelength
contributions in the post-singularity state.
The first occurs  as a direct
consequence of the sign change in (\ref{eq:matchb}),
and is independent of the amount of radiation generated
at the singularity. The second is proportional 
to the difference in the densities of the radiation
on the two branes. 
At leading order in velocities we have 
\be
\Delta \zeta_{4,M} = 
 \frac{3}{64}\frac{\epsilon_0}{k^2L^2}(V_{in}^4+V_{out}^4)
-\frac{(r_+-r_-) \epsilon_0 V_{out}^2}{32 k^2}
+O(r_{\pm} V^3, V^5L^{-2},
\rho_{\pm}^2 L^2), 
\ee
where $V_{in}$ and $V_{out}$ 
are the relative velocities of the branes before and
after collision. Note
that since $P \propto \epsilon_0$, matching $P$ is in  fact
equivalent to matching $\epsilon_0$ across
the collision as proposed in Ref. ~\ref{ekperts}. In
terms of four-dimensional
 parameters defined in Section III.C including $r_4$ given
in (\ref{eq:rdef}) 
defining the abundance of the radiation
and the fractional density mismatch $f$ defined in
(\ref{eq:fraction}), we find again
at leading order in velocities
\be
\Delta \zeta_{4,M} = 
\frac{3}{64}\frac{\epsilon_0}{k^2L^2}(V_{in}^4+V_{out}^4)
-\frac{3 \epsilon_0}{16 k^2} {f r_4 V_{out}^3\over L}.
\labeq{fin1}
\ee
This is our final result, relevant to tracking perturbations
across the singularity in the ekpyrotic and cyclic models.
We see it consists of two essentially independent terms. The first
is proportional the radiation density mismatch on the two
branes after collision. Note that just such a mismatch (with
more radiation on the negative tension brane) was required
in order to enable the cyclic solution of Ref.~\ref{STu}
to work. The second term exists however even in the limit
of no radiation generated on the branes. As we 
have noted above, it is nonzero even if $V_{in}=V_{out}$, and
it originates in 
the sign change of the parameter $Q$ in
our matching rule, which yields an arrow of time 
across the collision as explained in Section II. Going back
to the original formula (\ref{eq:match}) in which we
have not made the small velocity approximation, we
note that both the radiation-dependent and radiation-independent
terms possess a well defined limit as the
brane collision becomes relativistic (as the rapidity 
$y_0 \rightarrow \infty$),
\be
\Delta \zeta_{4,M} \approx
\frac{\epsilon_0}{k^2}\left({9\over 4 L^2} + {(r_+-r_-)\over 8}\right).
\labeq{finfast}
\ee
Recall, we need the radiation densities on the branes to be much
smaller than their tension i.e. 
$r_\pm L^2 \ll 1$, in order that
the four-dimensional effective theory be valid (Section III).
Therefore in the high velocity limit, the radiation-independent
term dominates. 
Conversely, from (\ref{eq:fin1}), in the low velocity limit 
(with $ (r_+-r_-)L^2$ fixed) the radiation-dependent 
term dominates.

We should stress once more that
the dependence upon parameters in (\ref{eq:fin1}) 
indicates its thoroughly five-dimensional
 origin. It cannot be expressed in 
purely four-dimensional
terms.
In previous work\cite{ekperts} with Khoury and Ovrut, two of us employed 
a more naive matching prescription framed 
entirely in terms of the four-dimensional effective theory.
This prescription was based upon using the comoving
energy density perturbation $\epsilon_m$, which is finite at the
singularity, as the matching variable. Unfortunately,
since the differential equation governing 
$\epsilon_m$ is singular at $t=0$, the  
first time derivative of $\epsilon_m$ is not an independent quantity at
the collision and hence could not be independently matched.
Instead we proposed matching the 
second time derivative. This has the virtue of at least yielding 
a dimensionally correct result, but it is ambiguous since there
are other choices of finite variables.
Now we understand the source of the ambiguity better. There
is simply not enough
information present in the four-dimensional theory to fix the gauge.
For that, the 
five-dimensional picture is essential as we have seen here.

In summary, we have found that
a spectrum of  scale invariant,  growing, 
long wavelength perturbations generally
propagate across the singularity even in the limit when
no radiation is produced. The radiation-independent
contribution rests upon the sign
change of $Q$ in the matching rule (\ref{eq:matcha}).
If radiation is produced at the bounce, 
then,  for the long wavelength
modes we are interested in, we believe it is reasonable to
model the production of radiation
as occurring suddenly, taking into account
the conservation of energy and momentum as was done in Ref.~\ref{STu}.
In this case, 
we find an additional contribution to the long wavelength
scale-invariant perturbations emerging from the singularity, which
is proportional to the difference in the radiation densities on the
two branes.

\section{Conclusions}

In this paper we have developed an unambiguous and,
we believe, compelling rule for matching perturbations
across the types of singularity encountered in the
ekpyrotic and cyclic Universe scenarios. In the simplest
realization of these scenarios, involving the collision
of two $Z_2$ branes in a bulk with a negative cosmological
constant, we have shown that the proposed rule leads
unambiguously to a
spectrum of scale invariant growing 
density perturbations in the ensuing
hot big bang phase,
even in the limit when only a small amount of radiation 
is produced at the collision.  The result provides 
support for a key assumption of the ekpyrotic and
cyclic models.

We have dealt here only with the linear theory,
treating the perturbations as free massless fields which we
match across the singularity.  This treatment
clearly is not fully 
consistent since the perturbations are 
divergent at the singularity and nonlinear effects must become
important there. However, there are reasons to expect that
in the nonlinear theory, a similar matching rule will
apply. In the linear theory, we have seen that
the metric components
typically behave as 
$1+\epsilon {\rm ln} |t| $ as $t \rightarrow 0$. This is 
just the small
$\epsilon$ expansion of $|t|^\epsilon$, the generic
behavior expected in the full nonlinear
Kasner solutions of general relativity that 
describe the generic approach to a space-like singularity. 
The natural extension of our proposal to the nonlinear
theory is,  therefore, that we should match the Kasner
exponents across the singularity. As in the linear theory,
the canonical momenta associated
with the three-metric are finite and our proposal 
amounts to matching them with a sign flip. But our matching 
proposal
in (\ref{eq:matcha}) also reverses the long
wavelength component of the constant term in the metric
perturbation. Generalizing to the nonlinear case, we
may anticipate that when the metric tends to the Kasner
form with spacelike components 
$\sim e^Q |t|^P$, with $Q$ and $P$ of order $\epsilon$,
these components will match to
$e^{-Q}|t|^P$ in the outgoing state. If  $\epsilon$
is small as expected in the ekpyrotic/cyclic
scenarios,
nonlinear corrections will be of order 
$\epsilon^2$ and hence negligible.

Strongly rupporting the idea of a local matching rule
 is the classic conjecture  that in general relativity
the behavior of the metric and fields becomes ultralocal
in the approach to this type of singularity.\cite{BKL}
That is, the spatial derivatives become unimportant
and  the geometry at each point in space
follows a homogeneous Friedmann-Robertson Walker 
evolution that just depends on local conditions.   One might worry 
that
 contraction also leads to  chaotic mixmaster behavior in which
 the universe moves 
from one kind of Kasner contraction to another  and the
Kasner exponents change unpredictably.
However, the existence of mixmaster behavior depends on the  
number and types of fields.  We discuss elsewhere\cite{erickson} 
how the mixmaster behavior  can 
be naturally avoided in ekpyrotic and cyclic models.


%
Finally, with a precise matching rule for propagating 
perturbations through the singularity in place,
we believe that the cyclic and ekpyrotic models are
now on firmer footing. A detailed study applying the above results
to these cosmological scenarios will be given 
elsewhere\cite{kst}.

\section{Acknowledgements}

We would like to thank Martin Bucher, 
Ruth Durrer,
Chris Gordon,
Steven Gratton, 
Stephen Hawking, 
Gustavo Niz,
Claudia de Rham, 
Carsten Van de Bruck
and Toby Wiseman for useful discussions.
Special thanks are due Justin Khoury who took time away from the 
Stanley Cup Finals to offer helpful comments.
AJT acknowledges the support of an EPSRC
studentship. The work of NT is supported by PPARC (UK).
PJS is supported by
US
Department of Energy grant DE-FG02-91ER40671.

\section*{Appendix 1: Projected Einstein equation}

In Section III we derived the four-dimensional effective action 
for solutions with
cosmological symmetry and then used general co-ordinate invariance to
infer the covariant moduli space action. While we have shown that this
approach recovers
the cosmological solutions perfectly at low densities, we have 
obtained the low energy effective action describing the general (asymmetrical) 
case 
by simply assuming locality
and imposing covariance. While this is plausible it is important
to check it explicitly. This has in fact been done in 
Refs. ~\ref{Wiseman},\ref{Soda},\ref{Shiromizu} which
further clarify the conditions under which the moduli approximation is
valid. We shall compare the results of these works with those
of the moduli space approach.

We shall first show that the effective theory we have derived
satisfies one non-trivial check.
One way of formulating a low energy theory 
for the brane geometries is the so called Gauss-Codazzi
formalism developed in Ref.~\ref{Shiromizu2}. 
Here we take the five-dimensional
Einstein equations
and project them
onto the brane to infer an equation for the brane geometry. One 
finds
\be
G_{\mu\nu}^{\pm}=\pm\frac{1}{M_5^3 L}
T_{\mu\nu}^{\pm}+\frac{1}{M_5^6}S_{\mu\nu}^{\pm}-E_{\mu\nu}^{\pm},
\ee
where $T_{\mu\nu}^{\pm}$ is the stress-energy on the brane, not
including the tension. This looks
like the four-dimensional  Einstein equations except for two additional source
terms. One contains stress energy squared terms,
\be
S_{\mu\nu}=\frac{1}{12}T\, T_{\mu\nu}-\frac{1}{4} T_{\mu\alpha}
T^{\alpha}_{\nu}+\frac{1}{24}g_{\mu\nu}(3T_{\alpha \beta}T^{\alpha \beta}-T^2),
\ee 
where $T=T^{\lambda}{}_{\lambda}$, and the second $E_{\mu\nu}^{\pm}$ is obtained
from projecting the `electric' part of the bulk Weyl tensor onto the brane
\be
E_{\mu\nu}=\frac{\partial x^A}{\partial x^{\mu}} \frac{\partial
x^B}{\partial x^{\nu}} E_{AB}, \qquad E_{AB}=C_{ACBD}n^Cn^D,
\ee
where $n^A$ is the normal to the brane. Note that by definition
$E_{\mu\nu}$ is symmetric. Since this term
contains information about the second `$y$' derivatives of the bulk
geometry we cannot calculate it in any purely four-dimensional way and so
although the above equations strongly resemble Einstein's equations
they are purely formal. However we can construct one purely four
dimensional equation because $E_{\mu\nu}$ does satisfy the exact condition
\be
E_{\pm}^{\mu}{}_{\mu}=0.
\ee
The moduli space approximation only works in the limit in which the
stress-energy of the matter on the brane is much smaller than the
brane tension. This amounts to neglecting the $T^2$ terms in the
above action leaving
\be
G_{\mu\nu}^{\pm}=\pm \frac{1}{M_4^2}T^{\pm}_{\mu\nu}-E_{\mu\nu}^{\pm}.
\ee
From now on we shall for convenience use units where $M_4=(8\pi
G)^{-1/2}$ is unity. As a consequence of the Bianchi identities it
follows that in this `low energy' approximation the following
condition must be true.
\be
\nabla_{\mu}E_{\pm}^{\mu}{}_{\nu}=0.
\ee
Since $E_{\mu\nu}$ is conserved and traceless it means that the
influence of the bulk on the brane geometry is identical in form to that of
the stress energy of a conformal field theory. If we look for a
cosmological solution, the vanishing trace condition
tells us that the only non-zero components of $E^{\mu}_{\nu}$ are,
$E^0_0=f(b)$ and $E^i_j=-\frac{1}{3}f(b)\delta^i_j$ where $f(b)$ is
an arbitrary function of the scale factor on the brane. In addition
the condition that $\nabla_{\mu}E_{\pm}^{\mu}{}_{\nu}=0$ tells us that
$f(b)=C/b^4$ and so the effect of this term is gravitationally indistinguishable from radiation, and 
it may be thought of as a dark radiation term. This is the import of
Birkhoff's theorem in the bulk, viewed from the brane.

\smallskip

The moduli space approximation as we have developed it
 provides a precise prediction for
$E_{\mu\nu}^{\pm}$. A non-trivial check on this approximation is that
the predicted value of $E_{\mu\nu}^{\pm}$ is traceless. This condition
of tracelessness is built in at the start in the other formalisms
\cite{Wiseman,Soda,Shiromizu}, but is a nontrivial check
of our approach. 
We can compute the trace
by simply conformally transforming the trace of the Einstein
equation in the four-dimensional effective theory. Writing the brane metrics as
$g_{\mu\nu}^{\pm} dx^{\mu} dx^{\nu}=\Omega_{\pm}^2 g_{\mu\nu}dx^{\mu}
dx^{\nu}$ we find
\begin{eqnarray}
	\nonumber & &
	 E^{\mu}{}_{\mu}=-G^{\mu}_{\pm}{}_{\mu} \pm T_{\pm}=R_{\pm}  \pm
	 T_{\pm} \\
	\nonumber & &
	=\Omega_{\pm}^{-2}(R-\frac{6}{\Omega_{\pm}} \nabla^2
	 \Omega_{\pm})\pm T_{\pm} \\
	\nonumber & &
	=\Omega_{\pm}^{-2}(-T_4+(\nabla \phi)^2
	(1-6\frac{(\Omega_{\pm})_{,\phi\phi}}{\Omega_{\pm}})
	-\frac{6(\Omega_{\pm})_{,\phi}}{\Omega_{\pm}}
	 \nabla^2 \phi) \pm T_{\pm} \\
	& &
	=\Omega_{\pm}^{-2}(-T_4
	-\frac{6(\Omega_{\pm})_{,\phi}}{\Omega_{\pm}}
	 \nabla^2 \phi) \pm T_{\pm},
\end{eqnarray}
where $T_{\pm}=T^{\mu}_{\pm}{}_{\mu}$ and in the last step we have 
used $\Omega_+=\cosh(\phi/\sqrt{6})$, $\Omega_-=-\sinh(\phi/\sqrt{6})$.
Finally, making use of the equation of motion for the scalar field
\be
\nabla^2 \phi=-\frac{1}{4}(\Omega_+^4)_{,\phi} T^+-\frac{1}{4}(\Omega_-^4)_{,\phi} T^-,
\ee
we find that
\be
E_{\pm}^{\mu}{}_{\mu}=0.
\ee
It is interesting to note that 
the intermediate steps in this calculation {\it require}
that the conformal 
factors on the positive and negative tension branes are of
the forms described above involving $\cosh(\phi/\sqrt{6})$ or
$\sinh (\phi/\sqrt{6})$.

In order to compute the projected Weyl curvature
in general it is helpful to work at the level of the action. We start
with the action for the four-dimensional effective theory
\be
S=\int d^4x \sqrt{-g} (\frac{1}{2} (R-(\nabla \phi)^2) +\Omega_+^4 {\mathcal
L_+}+\Omega_-^4 {\mathcal L_-}).
\ee
To get the action for the metric on the positive tension brane we
simply perform the conformal transformation, taking us out of
Einstein frame
\be
S=\int d^4x \sqrt{-g_+} \Omega_+^{-4} (\frac{\Omega_+^2}{2} (R_+-6\Omega_+
\nabla^2_+ \Omega_+^{-1}-(\nabla_+ \phi)^2)
+\Omega_+^4 {\mathcal
L_+}+\Omega_-^4{\mathcal L_-}),
\ee
then defining $\Psi= \Omega_+^{-2}$ and performing an integration by
parts we obtain the following action for the metric on the positive
tension brane
\be
S_+=\int d^4x \sqrt{-g_+} (\frac{1}{2} (\Psi R_+-\frac{3}{2(1-\Psi)}
(\nabla_+ \Psi)^2 ) +{\mathcal
L_+}+(1-\Psi)^2 {\mathcal L_-}),
\ee
and a similar calculation on the negative tension brane defining
$\Phi=\Omega_-^{-2}$ gives
\be
S_-=\int d^4x \sqrt{-g_-} (\frac{1}{2} (\Phi R_-+\frac{3}{2(1+\Phi)}
(\nabla_- \Phi)^2 ) +{\mathcal
L_-}+(1+\Phi)^2 {\mathcal L_-}).
\ee
These results are in perfect agreement with the low energy
approximation developed in Refs.~\ref{Wiseman},\ref{Soda} using a metric
based approach and in Ref. ~\ref{Shiromizu} using the covariant curvature formalism.
After deriving the equations of motion by varying these actions we can simply read off the predictions for
the projected Weyl tensor on the positive tension brane as
\begin{eqnarray}
\nonumber & &
E^{\mu}_+{}_{\nu}=T^{\mu}_+{}_{\nu}(1-\frac{1}{\Psi})-\frac{(1-\Psi)^2}{\Psi}T^{\mu}_-{}_{\nu}
\\
\nonumber & &
-\frac{1}{\Psi}(\nabla_+^{\mu} \nabla^+_{\nu}\Psi -\delta^{\mu}_{\nu}
\nabla_+^2 \Psi) \\
& &
-\frac{3}{2} \frac{1}{\Psi (1-\Psi)}(\nabla^{\mu}_+ \Psi
\nabla_{\nu}^+ \Psi -\frac{1}{2} \delta^{\mu}_{\nu} (\nabla_+ \Psi)^2),
\end{eqnarray}
and on the negative tension brane
\begin{eqnarray}
\nonumber & &
E^{\mu}_-{}_{\nu}=-T^{\mu}_-{}_{\nu}(1+\frac{1}{\Phi})-\frac{(1+\Phi)^2}{\Phi}T^{\mu}_+{}_{\nu}
\\
\nonumber & &
-\frac{1}{\Phi}(\nabla_-^{\mu} \nabla^-_{\nu}\Phi -\delta^{\mu}_{\nu}
\nabla_-^2 \Phi) \\
& &
+\frac{3}{2} \frac{1}{\Phi (1+\Phi)}(\nabla^{\mu}_- \Phi
\nabla_{\nu}^- \Phi -\frac{1}{2} \delta^{\mu}_{\nu} (\nabla_- \Phi)^2)
\, .
\labeq{array}
\end{eqnarray}
A specially interesting limit of these equations is obtained by
$\phi \rightarrow 0$ implying $\Psi \rightarrow 1$ and $\Phi \rightarrow
\infty$ which corresponds to the distance between the branes becoming infinite. Providing we can neglect the derivative terms we see that in
this limit, matter on the positive tension brane couples to the brane
geometry by means of the conventional four-dimensional 
 Einstein equations, whereas
the geometry on the negative tension brane is dominated by its
coupling to matter on the positive tension brane, and will only 
start to look like conventional Einstein gravity
if a `stabilization' mechanism exists which freezes $\phi$ to a
constant value. In the latter case,
stress energy on each brane acts like a dark matter
source for gravity on the other brane.   

These equations (\ref{eq:array}) describe matter 
interacting in an unconventional way with gravity,
and yield a more complicated perturbation theory than usual.
Our approach makes it clear that it is simpler to work with
the effective four-dimensional theory in Einstein frame with a scalar field with a
canonical kinetic term, and then simply to use the map
$g^+_{\mu\nu}=(\cosh(\phi/\sqrt{6}))^2 g_{\mu\nu}$ and
$g^-_{\mu\nu}=(-\sinh(\phi/\sqrt{6}))^2 g_{\mu\nu}$ to infer the brane
geometries. The only sense in which this theory differs from
conventional four-dimensional physics is that the different forms of matter couple
non-minimally to gravity though the scalar field. 

\section*{Appendix 2: Gauge Invariant Variables}

As in four-dimensions the cosmological symmetry of the background metric
allows us to find a set of gauge invariant variables, which facilitates
the comparison of two different gauges. What the natural gauge invariant
variables are depends on the form of the background and our definition
closely follows but are not identical to those in Ref. \ref{Carsten}.

\smallskip

We begin with the background metric written in the form
\be
ds^2=n^2(t,y)(-dt^2+t^2 dy^2)+b^2(t,y) \delta_{ij} dx^idx^j.
\ee
We shall only consider spatially flat cosmologies for simplicity but
the generalization to closed and open universes is easy. The most
general scalar metric perturbation can be written as
\begin{eqnarray}
	\nonumber & &
	ds^2=n^2 (-(1+2\Phi)dt^2-2Wdtdy+t^2(1-2\Gamma)dy^2)  \\
	\nonumber & &
	-2 \nabla_i\alpha dx^idt+2t^2 \nabla_i \beta dydx^i \\
	 & &
	+b^2((1-2\Psi) \delta_{ij}-2 \nabla_i \nabla_j \chi) dx^idx^j ,
\end{eqnarray}
writing the perturbed metric as $g_{AB}+h_{AB}$ where $g_{AB}$ is the
background metric, then under a gauge transformation $x^A \rightarrow x^A+\xi^A$
the metric perturbation transforms as
\be
h_{AB} \rightarrow h_{AB}-g_{AC} \partial_B\xi^C-g_{BC} \partial_A\xi^C-\xi^C
\partial_C g_{AB}.
\ee
Since a five-vector $\xi_A$ has three scalar degrees of freedom $\xi^t$,
$\xi^y$ and $\xi^i=\nabla_i \xi^{s}$,  only four of the seven functions
$(\Phi,\Gamma,W,\alpha,\beta,\Psi,\chi)$ are physical. This immediately
tells us that we expect to be able to define four  gauge invariant
variables constructed from the metric alone. Let $\dot{A}$ denote
$\frac{\partial A}{\partial t}$ and $A'$ denote $\frac{\partial
A}{\partial y}$. Under a gauge transformation each of the variables
transforms as
\begin{eqnarray}
	\nonumber & &
	\Phi \rightarrow \Phi-\dot{\xi}^t-\xi^t \frac{\dot{n}}{n}-\xi^y \frac{n'}{n}, \\
	\nonumber & &
	\Gamma \rightarrow \Gamma+\xi'^y+\frac{1}{t}\xi^t+\xi^t 
	\frac{\dot{n}}{n}+\xi^y \frac{n'}{n}, \\
	\nonumber & &
	W \rightarrow W-\xi'^t+t^2 \dot{\xi}^y,  \\
	\nonumber & &
	\alpha \rightarrow \alpha-\xi^t+\frac{b^2}{n^2} \dot{\xi}^s, \\
	\nonumber & &
	\beta \rightarrow \beta-\xi^y-\frac{b^2}{n^2 t^2} \xi'^s,  \\
	\nonumber & &
	\Psi \rightarrow \Psi+\xi^t \frac{\dot{b}}{b}+\xi^y \frac{b'}{b}, \\
	\nonumber  & &
	\chi \rightarrow \chi+\xi^s . \\
\end{eqnarray}
It is then relatively easy to construct the following gauge invariant
quantities
\begin{eqnarray}
	\nonumber & &
	\Phi_{inv}=\Phi-\dot{\tilde{\alpha}}-\tilde{\alpha} \frac{\dot{n}}{n}-\tilde{\beta} \frac{n'}{n}, \\
	\nonumber & &
	\Gamma_{inv}=\Gamma+\tilde{\beta}'+\frac{1}{t}\tilde{\alpha}
	+\tilde{\alpha} \frac{\dot{n}}{n}+\tilde{\beta} \frac{n'}{n}, \\
	\nonumber & &
	W_{inv}=W-\tilde{\alpha}'+t^2 \dot{\tilde{\beta}}, \\
  	\nonumber & &
	\Psi_{inv}=\Psi+\frac{\dot{b}}{b}\tilde{\alpha}+\frac{b'}{b}
	\tilde{\beta},
\labeq{gages}
\end{eqnarray}
where $\tilde{\alpha}=\alpha-\frac{b^2}{n^2} \dot{\chi}$ and
$\tilde{\beta}=\beta+\frac{b^2}{n^2 t^2} \chi'$. We then see that
there is a special gauge defined by $\chi=\alpha=\beta=0$ in which
\begin{eqnarray}
	\nonumber & &
	\Phi_{inv}=\Phi, \\
	\nonumber & &
	\Gamma_{inv}=\Gamma, \\
	\nonumber & &
	W_{inv}=W, \\
  	\nonumber & &
	\Psi_{inv}=\Psi.
\end{eqnarray}
We define this to be five-dimensional longitudinal gauge and so we see that the
gauge invariant variables equal the values of the metric perturbations
in longitudinal gauge, in perfect analogy with four-dimensional cosmological
perturbation theory. This gauge is characterized by being spatially
isotropic in the $x^i$ co-ordinates but in general there will be a
non-zero $t-y$ component of the metric.

\subsection*{Position of branes}

In general, the locations in $y$ of the perturbed branes will be different
in different gauges, and it is very important to understand this
location in each case. 
Remarkably, in
the case where the the brane matter has no anisotropic stress this is
 easy to establish. Start in the gauge $\alpha=\chi=0$. From
the above transformation rules we can see that we can always go to
this gauge using only $\xi^t$ and $\xi^s$ transformation. This then
leaves us with the freedom 
to perform any $\xi^y$ transformation 
such that the position of each brane
remains unperturbed. Then working out the Israel matching
conditions we find that $\beta$ on the branes is related to the
anisotropic part of the brane's stress energy. So if we are considering
only perfect fluids, for which the shear vanishes, 
then the Israel matching
condition gives $\beta(y=y_{\pm})=0$. We can then go to longitudinal
gauge $(\alpha=\beta=\chi=0)$ with the transformation $\xi^y=\beta$
alone. But since $\beta$ vanishes on the branes, so does $\xi^y$
implying that the brane trajectories are unperturbed. So we see that
for the special case of matter with no anisotropic stress the
locations of the branes in longitudinal gauge are their unperturbed
values $y=y_{\pm}$. We can then infer the position of the branes in an
arbitrary gauge by
means of the above gauge transformations to be
\be
y=y_{\pm}-\tilde{\beta},
\ee 
where $y_\pm$ are the background values.
In particular in the class of Milne gauges we have defined
in (\ref{eq:gauge}) the branes are located at
\be
y=y_{\pm}-\frac{q^2}{t^2} \chi'.
\ee

\section*{Appendix 3: Birkhoff's Theorem and the background metric}

The bulk geometry considered in this paper solves 
the five-dimensional Einstein's equations sourced
by a pure negative
cosmological constant. For the background solution we restrict
to solutions possessing cosmological 
symmetry on three dimensional spatial slices. In close analogy
to the familiar situation for spherical symmetry in $3+1$ dimensions,
a Birkhoff-type theorem guarantees that in our case that away
from the branes, the 
background
must take the form of
either Anti-de Sitter (AdS) space-time, Schwarshild-AdS or AdS with a naked
singularity. In each case the metric may be written as
\be
ds^2=(\frac{r^2}{L^2}+k-\frac{\mu}{r^2})^{-1} dr^2
-(\frac{r^2}{L^2}+k-\frac{\mu}{r^2})dT^2+r^2 \gamma_{ij} dx^i dx^j,
\labeq{a31}
\ee
where $\mu$ is the mass of the black hole, $\gamma_{ij}$ is the 
canonical metric on $S^3$, $H^3$ or $E^3$, with  
$k$ the corresponding spatial
curvature, and $L$ is the AdS radius defined by $\Lambda = -6 M_5^3/L^2$
with $M_5$ the five-dimensional Planck mass.
We are most interested in the case $k=0$, for which 
it is
useful to change variables from $r$ to $Y$ obtained by setting 
the first term in (\ref{eq:a31}) to equal $dY^2$, obtaining
\be
ds^2=dY^2-N(Y)^2 dT^2+A(Y)^2 d\vec{x}^2,
\labeq{bf}
\ee
where for AdS 
\be
A(Y)^2=N(Y)^2=\exp[2Y/L],
\ee
for Schwarzschild-AdS with a horizon at $Y=0$
\be
A(Y)^2=\cosh(2Y/L) \, {\rm and} \,  \qquad N(Y)^2 =\frac{\sinh(2Y/L)^2}{\cosh(2Y/L)},
\labeq{sads}
\ee
and for AdS with a naked singularity at $Y=0$
\be
A(Y)^2=\sinh(2Y/L), \qquad N(Y)^2 =\frac{\cosh(2Y/L)^2}{\sinh(2Y/L)}.
\ee

For any configuration of branes possessing cosmological
symmetry, even if the branes move the Birkhoff theorem
guarantees that the bulk geometry 
takes one of the three forms above\cite{gregory,martinneil}.
In our case,
where the branes are $Z_2$-symmetric and have their 
tensions tuned to allow static empty brane
solutions, 
the only bulk solution that is consistent with moving branes
is the Schwarzshild-AdS solution. Consequently this is the
background five-dimensional metric we use in this paper. 

Technically, in order to study the perturbations it is
much simpler if one changes coordinates to those
in which the branes are static and the
bulk is time-dependent. That it
is always 
possible to choose such a coordinate
system may be seen as follows. 
Start with the Birkhoff-frame metric (\ref{eq:bf}) with 
$A$ and $N$ given by (\ref{eq:sads}). First, 
 change variables from $Y$ to $Z$ defined by $dZ=dY/N$, with
$Z$ chosen to be zero at the collision event, so that 
\be
ds^2=N^2 (-dT^2+dZ^2)+A^2 d\vec{x}^2,
\ee
where $N$ and $A$ are now functions of $Z$. Defining lightcone
co-ordinates $T_{\pm}=T \pm Z$ we have
\be
ds^2=N^2 (-dT_+dT_-)+A^2 d\vec{x}^2.
\ee
We now recognize that the form of this metric is invariant under the
light-cone coordinate transformation, 
$\tau \pm y=f_\pm(T\pm Z)$, 
which takes the metric to the form
\be
ds^2={N^2\over f'_+f'_-}(-d\tau^2 +dy^2) +A^2 d\vec{x}^2.
\ee
Now we set $t=\pm e^{\pm \tau}$, to describe the 
post- or pre-collision space-times respectively, and 
define $t^2 n^2(t,y)= N^2/(f'_+f'_-)$ and $b^2(t,y)=A^2$ to obtain 
\be
ds^2=n^2(t,y)(-dt^2+t^2 dy^2)+b^2(t,y)d\vec{x}^2,
\ee
which is the form used in this paper.

We now show that we can always choose the functions $f_\pm$ 
to make the branes static in the new coordinates. To see
this note that the new spatial coordinate
\be
y(T,Z)={1\over 2} \left(f_+(T+Z)-f_-(T-Z)\right)
\labeq{yeq}
\ee
itself satisfies the massless field equation in two dimensions.
If the two brane trajectories are  $Z=Z_\pm(T)$ in the $T,Z$ 
coordinates, then it follows from
the general theory of the wave equation that we
can always solve (\ref{eq:yeq}) for arbitrary chosen $y(T,Z)$
on two specified timelike curves $Z=Z_\pm(T)$. In particular we are
free to choose constant values 
$y=y_+$ on the positive 
tension brane and $y=y_-$ on the negative tension brane. 
Even after this choice there is additional coordinate freedom,
since to determine the solution for 
$y(T,Z)$ we need to specify additional Cauchy data, for 
example on a $T=$constant surface.

In practice we find it is straightforward to solve these
equations as a power series in $t$. 
The Israel matching conditions on the two
branes in Birkhoff coordinates read
\be
\tanh (2 Y_\pm/L)=(1\pm {\rho_\pm L^2\over 6})\sqrt{1-N^{-2}(Y_\pm) (dY_\pm/dT)^2},
\labeq{iseq}
\ee
where $\rho_\pm$ are the densities of matter or radiation on the branes.
In our case, when only radiation is present, and
we normalize the brane scale factors to be unity at collision
(cf. Section III.C), 
we have $\rho_\pm = r_\pm /A^4(Y_\pm)$. 
Equation (\ref{eq:iseq}) is a first order differential 
equation for the brane trajectories $Y_\pm(T)$, allowing them
to be 
straightforwardly determined 
as Taylor series in $T$.
Likewise we may solve 
explicitly for $Z$,
\be
Z(Y)={L\over 2} \left({\rm tan}^{-1} (x) +{1\over 2} \ln({x-1\over x+1})\right),
\labeq{zsol}
\ee
where $x^2\equiv \cosh(2Y/L)$, and hence obtain $Z_\pm(T)$
as a Taylor series in $T$. From (\ref{eq:yeq}) we obtain
\be
y_\pm = {1\over 2} \left(f_+(T+Z_\pm(T))-f_-(T-Z_\pm(T))\right),
\labeq{ypmeq}
\ee
which we may differentiate with respect to $T$, noting that
the $y_\pm$ are  constant, to obtain
\be
f_+'(T+Z_\pm(T))(1+V_\pm(T)) = f_-'(T-Z_\pm(T))(1-V_\pm(T)),
\labeq{fgeq}
\ee
where $V_\pm(T) \equiv (d Z_\pm(T)/dT)$ are the brane velocities. These two
equations may be simultaneously solved 
as a power series in $T$ with the
ansatz $f_\pm'(z)= z^{-1}+f^0_\pm +f^1_\pm z +\dots$. They
 are both trivially satisfied at
order $T^{-1}$. At each subsequent power $T^n,
n\geq 0$ one obtains two equations which fix the two constants
$f_\pm^n$. Finally, writing $f_\pm(z)=c_\pm + 
\ln z +f^0_\pm z  +f^1_\pm z^2/2 +\dots$, with $c_\pm$ 
constants, we can write the equation for $y_\pm$ and take the
limit $T\rightarrow 0$ on the right hand side  to obtain
\be
y_\pm= {1\over 2} (c_+-c_-)+{1\over 2} \ln \left({1+V_\pm \over 1+V_\pm}\right)
= {1\over 2} (c_+-c_-)+\theta^B_\pm,
\labeq{ypmneq}
\ee
where $\theta^B_\pm$ are the rapidities of the positive and negative
tension branes in the Birkhoff frame. Likewise we
obtain (for $t>0$)
\be
\tau={1\over 2} (c_++c_-)+\ln t.
\labeq{tpeq}
\ee
Setting $t=\pm e^{\pm \tau}$ as we do for $t>0$ or $t<0$ respectively,
and choosing $y_+=-y_-$ (i.e. the Lorentz frame in 
which the branes have equal and opposite speeds), then fixes $c_+=-c_-=
-{1\over 2} (\theta^B_++\theta^B_-)\equiv - \overline{\theta^B}$.
Now one may invert the equations $\tau \pm y=f_\pm(T\pm Z)$
to express $T+Z$ as a Taylor series 
in $t e^y$ for $t>0$
(or $t e^{-y}$ for $t<0$) and 
similarly $T-Z$ as a Taylor series in $t e^{-y}$
(or $t e^y$).
For example, post-collision one obtains
\be
T+Z= t e^{y}e^{\overline{\theta^B}}+O(t^2), \qquad
T-Z=t e^{-y}e^{-\overline{\theta^B}}+O(t^2),
\labeq{tpmz}
\ee
equations which will be useful in Appendix 4.
Hence we 
completely determine the metric functions $n^2$ and $b^2$ 
as Taylor series in $t e^{y}$ and $t e^{-y}$.
Finally, by rescaling
$t$ and $\vec{x}$ we can also ensure that in the new 
coordinates, 
$n(t,y)=1+O(t)$ and $b(t,y) =1+O(t)$.

\smallskip

As a check of this procedure, or indeed an alternative to it, one can 
directly solve Einstein's equations in the
frame in which the branes are static. The extrinsic curvature is given
by
\be
K_{\mu\nu}dx^{\mu} dx^{\nu}=\frac{1}{2nt}\partial_y g_{\mu\nu}dx^{\mu} dx^{\nu}=\frac{1}{2nt}(-(n^2)'dt^2+(b^2)'d\vec{x}^2),
\ee
and so the Israel matching conditions
\be
K_{\mu\nu}=\frac{1}{2 M_5^3}(T_{\mu\nu}-\frac{1}{3} g_{\mu\nu}T^{\lambda}_{\lambda}),
\ee
tell us that
\begin{eqnarray}
\frac{n'}{n^2t}=\frac{1}{L}\mp\frac{L}{3} \rho_{\pm} \mp \frac{L}{2} p_{\pm}, \\
\frac{b'}{nt b}=\frac{1}{L}\pm\frac{L}{6} \rho_{\pm}.
\labeq{israeli}
\end{eqnarray}
For the purposes of our analysis it will be convenient to define the
Lorentz frame we work in to be that in which the $y$ coordinates
of the branes (their
rapidities) are $y_\pm=\pm y_0/2$. Recall,
we also define the parameters $r_\pm$ to be the
the densities of radiation on each brane $\rho_{\pm}$ at collision,
and we treat these as free parameters. 
Through a direct series solution of the five-dimensional
Einstein equations, imposing the Israel matching conditions
(\ref{eq:israeli}) at each order in $t$, we obtain the
following solution for the background geometry
 near $t=0$:
\begin{eqnarray}
\nonumber & &
b(t,y)=1+(b_1 \sinh y+b_2 \cosh y)t +(e_0+e_1 \sinh 2y +e_2 \cosh
2y)t^2/2, \\
& &
n(t,y)=1+(d_1 \sinh y+d_2 \cosh y)t +(k_0+k_1 \sinh 2y +k_2 \cosh
2y)t^2/2, 
\end{eqnarray}
where the constant parameters are given by
\begin{eqnarray}
\nonumber & &
b_1=\frac{(12+L^2 (r_+-r_-))}{12L}{\rm sech} (y_0/2), \\
\nonumber & &
b_2=\frac{L}{12}(r_++r_-){\rm cosech} (y_0/2),  \\
\nonumber & &
d_1=\frac{(4-L^2 (r_+-r_-))}{4L}{\rm sech} (y_0/2), \\
\nonumber & &
d_2=-\frac{L}{4}(r_++r_-){\rm cosech} (y_0/2),  \\
\nonumber & &
e_0= \frac{1}{36 L^2}((-6+L^2 r_-)^2 +(6+L^2 r_+)^2+2(-6+L^2
r_-)(6+L^2 r_+)\cosh y_0, \\
\nonumber & &
\qquad +36(\cosh 2y_0 -1)) ({\rm cosech}y_0)^2, \\
\nonumber & &
e_1= \frac{1}{12}(-4-L^2(r_+-r_-))(r_++r_-) {\rm cosech} (y_0), \\
\nonumber & &
e_2= -\frac{1}{12L^2}((24+4L^2(r_--r_+)+2 L^4 r_+r_-+(-24-4L^2
(r_--r_+), \\
\nonumber & &
\qquad +L^4 (r_-^4+r_+^4)) \cosh y_0) ({\rm cosech}y_0)^2,\\
\nonumber & &
k_0=\frac{1}{6L^2}((21+L^4(r_+^2+r_-^2)+2(-12+L^4 r_-r_+) \cosh y_0+3
\cosh 2y_0) ({\rm cosech}y_0)^2, \\
\nonumber & &
k_1=-\frac{1}{12}(r_++r_-){\rm cosech}y_0(5L^2 (r_--r_+) +12 {\rm sech} y_0),\\
& &
k_2=\frac{1}{12L^2}(-24+10L^4 r_+r_-+(24+5L^4(r_+^2+r_-^2))\cosh y_0) ({\rm cosech}y_0)^2.
\labeq{5dsolpars}
\end{eqnarray}

\section*{Appendix 4: Meaning of the constants $c_1$ and $c_2$}

The two arbitrary gauge constants $c_1$ and $c_2$ in 
Eq.(\ref{eq:chiseries}) parameterizing the violation of
the asymptotically Neumann boundary condition (\ref{eq:normal})
have a simple geometrical
interpretation: they describe the displacement 
of the collision event in the $T,Y$ plane.
Recall that in the Neumann gauge, discussed in 
Section V.E, 
the brane trajectories are unperturbed and are 
are described by the equations $y=y_\pm=$ constant.
If we now gauge transform to an asymptotically
Neumann gauge, in which the normal derivatives
$n^{-1} t^{-1}\chi'(t,y_\pm)$ deviate from zero
at order $t$ as in (\ref{eq:chiseries}), we see that
the gauge transformation from conformal
Newtonian gauge to the Milne gauge we are in involves a
divergent $y$ coordinate displacement of 
$\xi^y = -q^2 \chi'/t^2$, which tends to $-(c_1\sinh y+c_2 \cosh y)/t$
plus a finite part 
as $t$ tends to zero. 
If $c_1=c_2=0$, then the 
perturbation in the brane $y$ coordinates, $\xi^y(y_\pm)$ is finite.
The rapidities of the two branes are perturbed, but the
collision event itself is still simultaneous as in 
Neumann gauge.  

In the remainder of this Appendix we provide a geometrical
interpretation of the two constants 
$c_1$ and $c_2$, showing that they parameterize the displacement
of the brane collision event away from its background location in
the embedding coordinates $T,Y$, 
at each $\vec{x}$.

If we start from Neumann gauge with $c_1=c_2=0$, we may 
introduce $c_1$ and $c_2$ via the following gauge transformation,
\begin{eqnarray}
\nonumber & &
\xi^s= (c_1 \cosh y +c_2 \sinh y)t, \\
\nonumber & &
\xi^y=-\frac{1}{t}(c_1 \sinh y+c_2 \cosh y),\cr
\nonumber & &
\xi^t=(c_1 \cosh
y+c_2 \sinh y).
\labeq{cor}
\end{eqnarray}
This is part of the gauge freedom described by the solutions to eq.(\ref{eq:g1})
and (\ref{eq:g2}). Although $\xi^y$ diverges near $t=0$, this is
merely a reflection of the singular nature of the Milne ($t,y$)
coordinate
system. 
In terms of the Birkhoff frame $T,Y$ coordinates defined in Appendix 3,
we find
\ba
\nonumber
\delta T &=& \frac{\partial T}{\partial t} \xi^t +\frac{\partial
T}{\partial y} \xi^y, \cr
\delta Y &=& \frac{\partial Y}{\partial t} \xi^t +\frac{\partial
Y}{\partial y} \xi^y. 
\labeq{disps}
\ea
Then using (\ref{eq:tpmz}) given in Appendix 2 
and  (\ref{eq:cor}) one infers
the displacement of the collision event 
\ba
\delta T&=& \left(c_1 \cosh \overline{\theta^B} -c_2 \sinh
\overline{\theta^B}\right),\cr
\delta Y&=& N(Y_c)\left(c_1 \sinh \overline{\theta^B} -c_2 \cosh
\overline{\theta^B}\right),
\labeq{lorenty}
\ea
independent of $y$ and hence holding for both branes. 
Here $N(Y_C)$ is the value of the
lapse function (given in Appendix C) at the collision value
of $Y$ in the Birkhoff frame, and $\overline{\theta^B}$ is the
mean rapidity of the two branes in that frame.
Therefore all the gauge transformation (\ref{eq:cor}) does
is to move the collision event around by an arbitrary
finite displacement in the $T,Y$ plane.


\end{document}